\documentclass[journal]{IEEEtran}

\usepackage{cite}				
\usepackage{graphicx}	
\usepackage{amsmath}		
\usepackage{algorithmic}	
\usepackage{array}		%
\usepackage{dsfont}	
\ifCLASSOPTIONcompsoc	
\usepackage[caption=false,font=normalsize,labelfont=sf,textfont=sf]{subfig}
\else
\usepackage[caption=false,font=footnotesize]{subfig}
\fi
\usepackage{stfloats}
\usepackage{algorithm}	
\usepackage{color}		
\usepackage[dvipsnames]{xcolor}
\usepackage{bm}			
\usepackage{amsfonts}	
\usepackage{amssymb}	
\usepackage{booktabs}   
\usepackage{makecell}   
\usepackage[stable]{footmisc}
\usepackage{dsfont}	    

\newtheorem{property}{Property}

\newtheorem{problem}{Problem}
\begin{document}
	\allowdisplaybreaks 
	\title{Random Caching Design for Multi-User Multi-Antenna HetNets with Interference Nulling}
	\author{Tianming~Feng, Xuemai~Gu,~\IEEEmembership{Member,~IEEE,} and Ben~Liang,~\IEEEmembership{Fellow,~IEEE} 
		\thanks{T. Feng and X. Gu are with the School of Electronic and Information Engineering, Harbin Institute of Technology, Harbin 150001, China (e-mail: \{fengtianming, guxuemai\}@hit.edu.cn).}
		\thanks{B. Liang is with the Department of Electrical and Computer Engineering, University of Toronto, Toronto, ON M5S 1A1, Canada (e-mail: liang@ece.utoronto.ca).}
		\vspace{-2em}}
	
	\markboth{IEEE}%
	{Draft paper}
	
	\maketitle
	\begin{abstract}
		The strong interference suffered by users can be a severe problem in cache-enabled networks (CENs) due to the content-centric user association mechanism. To tackle this issue, multi-antenna technology may be employed for interference management. In this paper, we consider a user-centric interference nulling (IN) scheme in two-tier multi-user multi-antenna CEN, with a hybrid most-popular and random caching policy at macro base stations (MBSs) and small base stations (SBSs) to provide file diversity. 
		All the interfering SBSs within the IN range of a user are requested to suppress the interference at this user using zero-forcing beamforming. Using stochastic geometry analysis techniques, we derive a tractable expression for the area spectral efficiency (ASE). A lower bound on the ASE is also obtained, with which we then consider ASE maximization, by optimizing the caching policy and IN coefficient. To solve the resultant mixed integer programming problem, we design an alternating optimization algorithm to minimize the lower bound of the ASE. Our numerical results demonstrate that the proposed caching policy yields performance that is close to the optimum, and it outperforms several existing baselines.		
	\end{abstract}
	
	\begin{IEEEkeywords}
		Random caching, HetNets, ZFBF, interference nulling, stochastic geometry.
	\end{IEEEkeywords}
	
	\IEEEpeerreviewmaketitle
	
	\section{Introduction}
	\IEEEPARstart{H}{eterogeneous} networks (HetNets) provide an effective framework to meet the exponentially increasing data traffic demand caused by the explosive growth of mobile devices \cite{Bhushan2014,Andrews2014}. By densely deploying small base stations (SBSs) along with the existing macro base stations (MBSs), HetNets can boost spatial utilization and thus significantly improve the area spectral efficiency (ASE). However, base station (BS) densification and the consequent spectrum reuse in HetNets can cause heavy burden on backhaul links \cite{Ge2014} and strong inter-cell interference \cite{David2011}, which may become bottlenecks on network performance.
		
	To alleviate the backhaul load, an effective solution is to cache popular files at BSs \cite{Liu2016a}. The paradigm of cache-enabled networks (CENs) is inspired by the fact that a large part of the data traffic is caused by the duplicate downloads of popular files, so that pre-caching the popular files at BSs during off-peak time can significantly reduce the backhaul cost \cite{Bastug2014}. Moreover, equipping BSs with cache can further improve the network ASE in HetHets \cite{Liu2016a}. However, because of the limited cache size at BSs, the serving BS of a user may not be geographically its nearest BS, which may lead to strong interference at the user \cite{Cui2017a}. Therefore, appropriate interference management techniques are needed to enhance the quality of signals received by users.
	
	Deploying multiple antennas at each BS is a widely-adopted method to address the aforementioned interference problem. With multiple antennas, the link reliability can be ensured by providing spatial diversity or conducting interference nulling (IN) \cite{Larsson2014}. In addition, the ASE can be further improved by serving multiple users over the same time-frequency resource block (i.e., achieving space division multiple access (SDMA)) \cite{Mu-mimo2012}.
	
	Recently, several works have focused on performance analysis and caching policy design in multi-antenna CENs \cite{Liu2017a,Kuang2019,Kuang2019a,Jiang2019c,Xu2019c}. However, \cite{Liu2017a,Kuang2019,Kuang2019a
	} do not consider interference management. In \cite{Jiang2019c}, multiple antennas are equipped at each receiver to suppress the received interference in CENs with random caching. However, this receiver-side IN scheme may not be effective when the network is highly densified. In \cite{Xu2019c}, the BSs are grouped into disjoint clusters, and the intra-cluster interference is canceled by BS coordination. However, the number of BSs for interference coordination is fixed, which can be ineffective in practice where different users have different numbers of dominant interferers due to the irregular placement of BSs. Further discussion on related work is given in Section~\ref{section: MU Related Works}. Overall, while it is expected that deploying multiple antennas is an effective method for managing interference and improving the ASE of CENs, 1) there has been no formal analysis on the relation between the IN requests received by each BS and the files cached in the network; and 2) it remains an open problem how to jointly design random caching policy and IN to take advantage of the cache resource as well as the antenna resource to improve the network ASE.

	In this paper, we consider a two-tier cache-enabled multi-user multi-antenna HetNet, where both MBSs and SBSs are equipped with caches and multiple antennas. We study an IN scheme for the SBS tier, where a part of the spatial degrees-of-freedom (DoF) is used to serve multiple users per channel to improve the network ASE; whereas the remaining DoF is reserved for IN to suppress the dominant interference received by the users. Due to the random caching at SBSs and the adoption of SDMA, the pattern of interference received by each user is complex. Moreover, the IN scheme itself further complicates the interference distribution and the analysis of the ASE. In this context, our main contributions are as follows.
	
	$\bullet$ We analyze the ASE of the aforementioned two-tier cache-enabled multi-user multi-antenna HetNets with a hybrid caching policy and a user-centric IN scheme in the typical interference-limited scenario. Specifically, the most popular caching (MPC) method in the MBS tier and random caching in the SBS tier are used to provide file diversity. All the antennas at MBSs are used to serve multiple users to achieve SDMA, whereas part of the spatial DoF at SBSs is used to implement SDMA, and the remaining DoF is reserved for IN. Due to the high complexity of the ASE expression, we further derive a simple lower bound for it, which is in the form of a sum of linear-fractional functions of the caching probabilities.
	
	$\bullet$ We consider the ASE maximization problem by optimizing the caching policy and the IN coefficient. We separate the problem into two sub-problems, i.e., cache placement optimization and IN coefficient optimization. The first sub-problem is a complicated mixed integer programming problem. By exploiting some of its structural properties, we significantly reduce the computational complexity of the discrete part of this problem. By replacing the ASE with the aforementioned simple lower bound, the continuous part is transformed into a convex problem, which can be effectively solved using KKT conditions. Then, the second sub-problem is effectively solved using line search. Finally, by alternately solving these two sub-problems, we reach a stationary point for maximizing the ASE lower-bound.
	
	$\bullet$ Our simulation results reveal that the proposed solution yields performance that is close to optimal. Specifically, when the signal to interference and noise ratio (SINR) threshold $\tau$ is small, caching files according to the uniform distribution will achieve the highest ASE; whereas caching the most popular files is better when $\tau$ is large. In general, the proposed caching policy outperforms existing caching strategies. 	

	The rest of this paper is organized as follows. We present a literature survey in Section \ref{section: MU Related Works}. The system model is presented in Section \ref{section: MU System}. Section \ref{section: MU Performance} provides a performance analysis of the system, and Section \ref{section: MU Max} optimizes the random caching policy and IN coefficient. The numerical results are provided in Section \ref{section: MU Numerical}, and the conclusions are drawn in Section \ref{section: MU Conclusion}.
	
	\textit{Notations:} In this paper, vectors and matrices are denoted by blod-face lower-case (e.g., $\mathbf{h}$) and upper-case (e.g., $\mathbf{H}$) letters respectively. We use $\mathbf{I}_n$ to denote an $n\times n$ identity matrix. $(\cdot)^T$ and $(\cdot)^H$ denote transpose and Hermitian (or conjugate transpose), respectively. $\mathbf{H} ^{\dagger} = (\mathbf{H}^{H} \mathbf{H})^{-1} \mathbf{H}^{H}$ denotes the left pseudo-inverse of $\mathbf{H}$. $\mathbb{R}$ and $\mathbb{C}$ are used to denote the set of real numbers and complex numbers, respectively. The complex Gaussian distribution with mean $\mu$ and covariance $\sigma^2$ is denoted by $\mathcal{CN} (0, \sigma ^2) $. $\|\mathbf{h} \|$ denotes the Euclidean norm of the vector $\mathbf{h}$, and $\|\mathbf{H} \|_1$ denotes the $L_1$ induced norm of the matrix $\mathbf{H}$, i.e., $\|\mathbf{H} \|_1 = \max_{1\leq j \leq n} \sum_{i = 1}^{m} \left| h_{ij} \right| $ for $ \mathbf{H} \in \mathbb{R} ^{m\times n}$. $X \overset{d}{\sim} Y$ means that $X$ is distributed as $Y$. $\mathbb{P}[\cdot]$ denotes the probability, while $\mathbb{E}[\cdot]$ is the expectation. $\Gamma (k, \theta)$ denotes the Gamma distribution with shape parameter $k$ and scale parameter $\theta$; $\Gamma(\cdot)$ denotes the Gamma function; $\mathds{1}\{\cdot\}$ is the indicator function.

\section{Related Works}\label{section: MU Related Works}
	\subsection{Caching Policy Design}
	Caching policy design plays a pivotal role in reaping the benefit of caching. By carefully designing the caching policy, more files can be stored in the network to provide file diversity and thus ensure network performance. In \cite{Bastug2015,Tamoor-ul-Hassan2015a,Bharath2016}, the authors propose three basic caching policies, i.e., MPC, uniform distribution caching (UDC), and independent identical distribution caching (IIDC), respectively. However, these simple caching policies do not take full advantage of the benefit brought by caching. Therefore, the authors of more recent works endeavor to design optimal random caching policies to respectively maximize the hit probability \cite{Wen2017,Zhang2020,Yang2020b}, the successful transmission probability (STP) \cite{Cui2016,Yang2020d,Liu2017a}, the traffic offloading gain \cite{Xu2019c,Amer2020}, and the ASE \cite{Liu2017a,Kuang2019,Xu2019c}. When adopting a random caching policy, a BS will randomly decide whether to store a file according to its caching probability. It is likely that the serving BS of a user is not geographically its nearest BS. In this case, the interference caused by the nearer BSs will significantly degrade the quality of signal received by the user. Therefore, for CENs with a random caching policy, we need to further consider interference management.
	\subsection{Interference Nulling with Multiple Antennas}
	Multi-antenna communication is a widely-adopted approach to increase SINR. With multiple antennas equipped at each BS, not only the desired signals of users can be boosted, but more effective interference management techniques can be implemented \cite{Akoum2013,Li2015a,Cui2016a,Zhu2018b}. Specifically, the authors in \cite{Akoum2013} propose a cluster-based IN scheme, where all the BSs are grouped into disjoint clusters, and zero-forcing beamforming (ZFBF) is conducted at each BS to mitigate intra-cluster interference. However, the scheme is designed from the perspective of transmitters and does not directly consider each user's needs. To this end, a user-centric IN scheme is proposed for multi-antenna small cell (single-tier) networks in \cite{Li2015a}, where an IN range is set for each user based on its desired signal strength and interference level. The authors in \cite{Cui2016a} further extend the IN scheme in \cite{Li2015a} to HetNets. The IN schemes considered in \cite{Akoum2013,Li2015a,Cui2016a} utilize extra spatial DoF at BSs to eliminate interference at users, and data exchange between BSs is avoided. 
	The work in \cite{Zhu2018b} achieves an improvement of SINR from another perspective, i.e., the network MIMO system, where cooperative BSs jointly transmit information to multiple users in the cooperative cluster via coherent beamforming. In this scheme, a BS does not need to reserve extra spatial DoF for IN, but the exchange of user data between cooperative BSs is required. We note that none of \cite{Akoum2013,Li2015a,Cui2016a,Zhu2018b} considers caching.
	\subsection{Multi-Antenna Cache-Enabled Networks}
	Random caching policy design in multi-antenna CENs has also been studied in recent literature. In \cite{Liu2017a}, an optimal caching policy is proposed to maximize the STP and ASE in a two-tier CEN, but only MBSs connecting to the core network are equipped with multiple antennas, while the SBSs with caches are equipped with only a single antenna. The authors in \cite{Kuang2019} consider a multi-tier multi-antenna CEN, where BSs from different tiers have different caching capabilities and are equipped with multiple antennas to serve multiple users via SDMA. A locally optimal caching policy for each tier is obtained by maximizing the potential throughput. The authors in \cite{Kuang2019a} design a locally optimal caching policy in a single-tier multi-antenna CEN considering limited backhaul capacity. The authors in \cite{Jiang2019c} attempt to perform interference cancellation at the receiver side in CENs. Two types of linear receivers at users are considered, i.e., maximal ratio combining receiver and partial zero-forcing receiver. Only the channel state information (CSI) at receivers is required, so that the burden on the BS can be reduced. In \cite{Xu2019c}, the authors investigate a BS coordination IN scheme for CENs, where a user-centric BS clustering model is proposed to form a BS cluster, and ZFBF is adopted at BSs to null out the interference within the coordination cluster. Optimal caching policies are obtained by maximizing the average fractional offloaded traffic and average ergodic spectral efficiency. 
		
	In contrast to these existing works, we focus on the analysis and maximization of the network ASE considering the proposed user-centric IN scheme. Different from \cite{Liu2017a,Kuang2019,Kuang2019a}, the relation between IN and caching is explored in this work. In contrast to \cite{Jiang2019c}, we consider the transmitter-side IN scheme, which is more effective since BSs usually have much more antenna resource than receivers. Moreover, unlike \cite{Xu2019c}, an adjustable IN coefficient is introduced to give the IN scheme more flexibility. Finally, in \cite{Jiang2019c,Xu2019c}, only one single user is served over each time-frequency resource block (RB), so the spatial DoF provided by multiple antennas is not fully utilized to realize SDMA. In this work, we consider the multi-user scenario, where SDMA is used to further improve the network capacity. All the aforementioned features make our system more complex and more challenging to analyze.
	\begin{table} [th]
		\centering
		\caption{Summary of Notation}
		\label{tbl: MU symbols}
		\begin{tabular}{|m{1.8cm}<{\centering}|m{0.68\columnwidth}|} 
			\hline
			\textbf{Notation} & \textbf{Description}\\
			\hline
			$\Phi_1$, $\Phi_2$, $\Phi_u$, $\Phi_u^2$ & PPPs of MBSs (named as the $1^{st}$ tier), SBSs (named as the $2^{nd}$ tier), users, and the users served by SBSs.\\
			\hline
			$\lambda_{1}$, $\lambda_{2}$, $\lambda_{u}$ & The densities of MBSs, SBSs, and users.\\
			\hline
			$M_k$ & Number of antennas equipped at BSs in the $k$-th tier.\\
			\hline
			$P_k$ & Transmit power of BSs in the $k$-th tier.\\
			\hline
			$U_k$ & Number of users served simultaneously by an BS from the $k$-th tier.\\
			\hline
			$\alpha_k$ & Path loss exponent for the $k$-th tier.\\
			\hline
			$\mathcal{N}$, $N$, $\mathcal{N}_1$, $N_1$, $\mathcal{N}_2$, $N_2$ &  Set of all files, size of $\mathcal{N}$, the first sub-set of files, size of $\mathcal{N}_1$, the second sub-set of files, size of $\mathcal{N}_2$.  \\			
			\hline
			$\mathcal{N}_c$, $N_c$,\qquad $\mathcal{N}_b$, $N_b$  & Set of files stored in the SBS tier, size of $\mathcal{N}_c$, set of files not stored in the network, size of $\mathcal{N}_b$.\\
			\hline
			$C_k$, $C_b$& Cache size of each BS in the $k$-th tier, backhaul capacity of each MBS.\\
			\hline
			$a_n$  & The popularity of file $n$.\\
			\hline
			$T_n$, $\mathbf{T}$ & Caching probability for file $n$, caching probability vector.\\
			\hline
			$\mathcal{I}$, $I$, $\mathcal{I}_i$, \quad $\mathcal{I}^n$, $I^n$ & Set of file combinations, size of $\mathcal{I}$, the $i$-th combination in $\mathcal{I}$, set of combinations containing file $n$, size of $\mathcal{I}^n$. \\
			\hline
			$p_i$ & Probability of an SBS storing $\mathcal{I}_i$.\\			
			\hline
			$x_{10}$, $x_{20}$ & Serving BSs of an MBS-user and SBS-user, respectively.\\	
			\hline
			$Z_{1}$, $Z_{2}$ & The distance between an MBS-user or SBS-user and its serving BS.\\	
			
			\hline
			$L_b$ & Backhaul load of an MBS.\\
			\hline		
			$\mu$ & The IN coefficient.\\
			\hline
			$\varTheta_x$, $\bar{\varTheta}$ & Number of IN requests received by an SBS $x$, mean number of IN received by an SBS.\\
			\hline
			$\Phi_2^a$, $\Phi_2^b$, $\Phi_2^c$, $\lambda_{2}^{a}$, $\lambda_{2}^{b}$, $\lambda_{2}^{c}$, $\varOmega_a$, $\varOmega_b$, $\varOmega_c$  & Sets of interfering SBSs of Type-A, Type-B, and Type-C; densities of $\Phi_2^a$, $\Phi_2^b$, $\Phi_2^c$; corresponding distance intervals of $\Phi_2^a$, $\Phi_2^b$, $\Phi_2^c$.\\
			\hline
			$\Upsilon_{n,k}$ & Received SINR of a user requesting file $n$ and served by the $k$-th tier.\\
			\hline
			$g_{0k}$, $g_{x_k}$ &  Desired channel gain in the $k$-th tier, interfering channel gain in the $k$-th tier.\\
			\hline
			$\tau$ & SINR threshold.\\
			\hline
			$\varepsilon$ & IN missing probability.\\
			\hline
		\end{tabular}
	\end{table}

\section{System Model}\label{section: MU System}
	\subsection{Network and Caching Model}
	We consider a two-tier cache-enabled multi-antenna HetNet, where a tier of SBSs is overlaid with a tier of MBSs. To avoid inter-tier interference, orthogonal frequencies are adopted at MBSs and
	SBSs. The locations of MBSs and SBSs are modeled as two independent homogeneous Poisson	point processes (PPPs) denoted by $\Phi_{1}$ and $\Phi_{2}$ with densities $\lambda_{1}$ and $\lambda_{2}$, respectively, where $\lambda_{1} < \lambda_{2}$. The users are also distributed in $\mathbb{R}^2$ according to a PPP $\Phi_{u}$ with density $\lambda_{u}$. Each BS in the $k$-th tier, $k \in \{ 1,2 \}$, is equipped with $M_k$ antennas, with transmit power $P_k$, and can serve $U_k$ users, where $U_k \leq M_k$, simultaneously over one time-frequency RB, i.e., intra-cell SDMA is considered. MBSs are equipped with caches and connected to the core network with limited-capacity backhaul links, whereas SBSs are only equipped with caches.
	Each user has a single receiving antenna. Full load assumption is considered in this paper, i.e., $ \lambda_{u} \gg \lambda_{k}, \forall k \in \{ 1,2 \} $, so that each BS in the $k$-th tier has at least $U_k$ users connected to it \cite{Li2016,Hattab2018,Gupta2014}. When the number of users in each cell from the $k$-th tier is greater than $U_k$, the BS will randomly choose $U_k$ users to serve at each time instant. We focus on the performance analysis of a typical user $u_0$ located at the origin without loss of generality. 
	
	We consider a content library containing $N$ different files denoted by $\mathcal{N} = \{1,2,\cdots,N\}$, all files have the same size which equals 1. At any given time instant, for any arbitrary given user, the probability that file $n$ is requested by the user is $a_n \in [0,1]$, which is called file popularity, satisfying $\sum_{n \in \mathcal{N}} a_n = 1$. For example, it has been observed that the popularity of file $n$ follows a Zipf distribution \cite{Liu2017a,Jiang2019c,Kuang2019}, i.e., $a_n = \frac{n^{-\gamma_z}}{\sum_{n\in \mathcal{N}} n^{-\gamma_z}}, \,\,n \in \mathcal{N}$, where $\gamma_z$ is the Zipf exponent. The analysis in this work is applicable to any popularity distribution. Without loss of generality, we assume that a file of smaller index has higher popularity, i.e., $a_1 > a_2 > \cdots > a_N $. Denote by $C_k$, $k \in \{1,2\}$ the cache size of each BS in the $k$-th tier, and $C_b$ the backhaul link capacity of each MBS.
	
	For the hybrid caching policy, we observe the following principles: 1) a high cache hit probability can be ensured by storing the most popular files at all MBSs, since MBSs are usually equipped with large cache space, the frequently requested files can always be served by the nearest MBS of a user; 2) spatial file diversity can be fully exploited by storing the less popular files at SBSs randomly, since the density of the SBS is comparatively high so that more files can be stored at the SBS tier; and 3) the requests for the files not stored at any BS can be satisfied via the backhaul links of MBSs. 
	
	Therefore, we divide the content library into two disjoint portions: the first sub-set of files contains the $N_1$ most popular files, which is denoted by $\mathcal{N}_1 \triangleq \{1,2,\cdots,N_1\}$; the second sub-set $\mathcal{N}_2 \triangleq \{N_1+1, N_1+2, \cdots,N\}$ contains the remaining $N_2 \triangleq |\mathcal{N}_2|$ files. For MBSs, the most popular caching policy is employed, where all the $N_1$ files from $\mathcal{N}_1$ are stored at every MBS; thus we explicitly require that $C_1 = N_1$. For SBSs, we adopt a random caching policy, where each SBS randomly chooses $C_2$ different files from $\mathcal{N}_2$ to store. Denote by $T_n \in [0,1]$ the probability that an SBS caches the file $n$. We further define  $\mathcal{N}_c \triangleq \{ n \in \mathcal{N}_2 : T_n >0 \}$, which is the set of files that can be stored in the SBS tier, with $N_c \triangleq |\mathcal{N}_c|$. Then, $\mathbf{T} \triangleq [T_n]_{n \in \mathcal{N}_c}$ denotes the caching probability vector for the SBS tier, which is identical for all the SBSs. Let $\mathcal{N}_b \triangleq \mathcal{N}_2 \backslash \mathcal{N}_c $ be the set of files not cached at any BS in the network, with $N_b \triangleq | \mathcal{N}_b| $. The backhaul link of each MBS is used to retrieve the files that are requested by its associated users but not stored at its local caches from the core network. Thus, the set of files needed to be retrieved from the core network of an MBS is a subset of $\mathcal{N}_b $.	
	
	With the random caching policy, each SBS stores $C_2$ different files out of $\mathcal{N} _c$. Thus, there are totally $I \triangleq \binom{N_c}{C_2}$ different combinations that can be chosen. Denote by $\mathcal{I} \triangleq \{ 1,2,\cdots,I \}$ this set of $I$ combinations. Let $\mathcal{I}_i$ be the $i$-th combination from $\mathcal{I}$. Let $\zeta_{i,n} = 1$ indicate that the file $n$ is included in combination $i$, and $\zeta_{i,n} = 0$ otherwise; thus $\sum_{n \in \mathcal{N}_c} \zeta_{i,n}  = C_2$. Let $\mathcal{I}^{n} \triangleq \{ i \in \mathcal{I}: \zeta_{i, n}=1 \}$ be the set of $I^n \triangleq \binom{ N_c -1}{ C_2-1}$ combinations containing file $n$. Let $p_i$ be the probability that an SBS chooses $\mathcal{I}_i$ to store; then we have
	\begin{equation}\label{equ: MU Tn and pi}
		T_n = \sum_{i \in \mathcal{I}^n } p_i, \quad n\in\mathcal{N}_c.
	\end{equation}

	\subsection{User Association}
	Based on the aforementioned hybrid caching policy, a user connects to a BS depending on the file it requests. Specifically, a user requesting file $n\in \mathcal{N}_1$ will connect to its nearest MBS; if the requested file $n$ of a user is stored at the SBS tier, i.e., $n \in \mathcal{N}_c$, the user will connect to its nearest SBS storing a combination $i \in \mathcal{I}^{n}$ (containing file $n$), and is referred to as an SBS-user; otherwise, a user will be served by its nearest MBS via the backhaul link. A user served by an MBS is referred to as an MBS-user. Denote by $x_{10}$ (resp. $x_{20}$) the location of the serving BS of a user if it is an MBS-user (resp. SBS-user). 
	
	In a multiuser MIMO system, when determining its potential connecting users, a BS normally uses its total transmit power to broadcast reference signals via a single antenna, and the multiuser beamforming is only conducted after user association \cite{Hattab2018,Li2016}. A user will choose a BS that can offer the maximum long-term average receive power for its requested file to connect. Note that for an MBS-user, the serving BS is its geographically nearest MBS. However, for an SBS-user, the serving BS may not be its nearest SBS, due to the random caching policy. This user association mechanism is referred to as \textit{content-centric association}, which is different from the distance-based association adopted in traditional networks \cite{Andrews2011,Gupta2014}. Therefore, the interference caused by the SBSs that are closer to the user than the serving SBS needs to be carefully handled. To this end, in this paper, we consider a \textit{user-centric interference nulling scheme}, which will be elaborated on later.

	Denote by $L_b$ the \textit{backhaul load} of an MBS, which is the number of different files requested by the users served by the backhaul link of the MBS. Due to the limited backhaul link capacity, each MBS can only retrieve at most $C_b$ different uncached files requested by its associated users from the core network. If $ L_b \leq C_b $, all the requested uncached files can be retrieved via the backhaul link; otherwise the MBS will uniformly randomly select $C_b$ different files to retrieve. The selected uncached files will firstly be retrieved from the core network by the MBS via its backhaul link, and then be passed on to the requesting users.	
	\begin{figure*}[!t]	
		\centering
		\includegraphics[height=6cm,width=17cm]{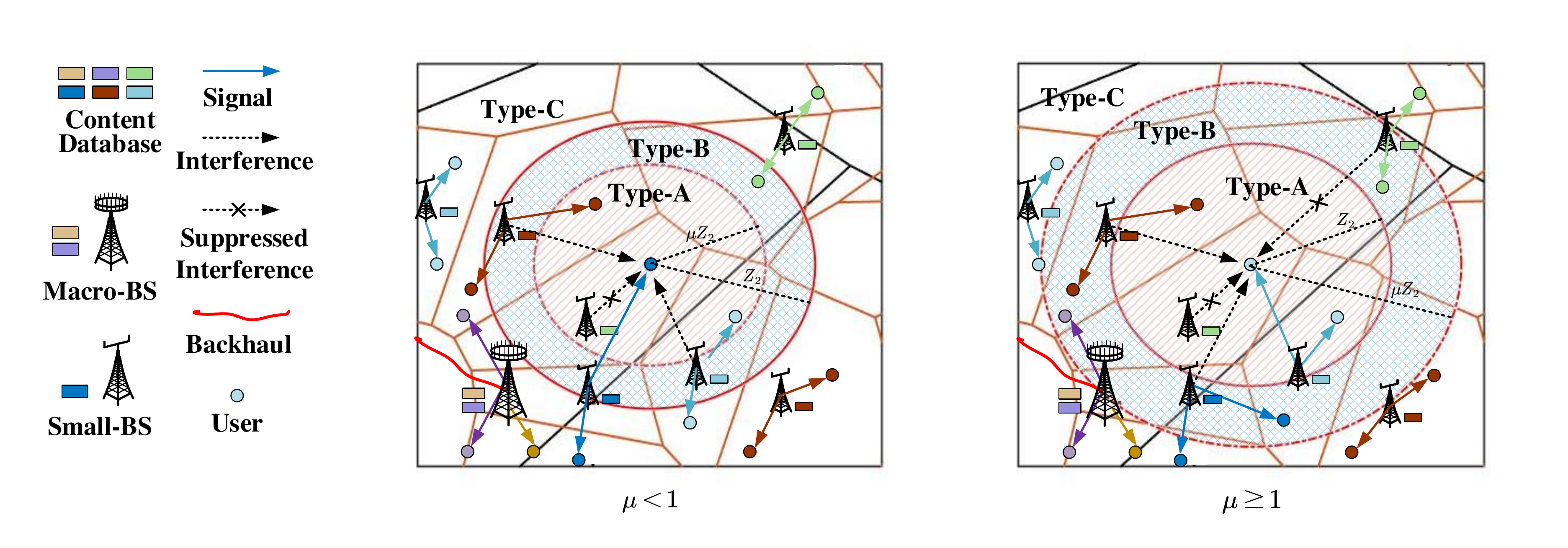}
		\caption{Network model. The two network tiers correspond to two Voronoi tessellations with black and brown solid lines, respectively. The typical user is located at the origin, the color of whom represents the file it requests. Three types of interfering SBSs are shown in the figure, and the interfering SBSs in the red dash circle will receive an IN request from the typical user.		  }
		\label{fig: MU system Model}
	\end{figure*}
	\subsection{Interference Nulling Scheme}\label{subsec: MU IN scheme}
	We will now elaborate on the user-centric interference nulling scheme used to suppress the interference from non-serving SBSs in the SBS tier. If the typical user $u_0$ requesting file $n$ is an SBS-user, the interference it receives mainly comes from 1) the SBSs that do not store file $n$, and are closer to $u_0$ than its serving SBS; 2) all SBSs that are further to $u_0$ than its serving SBS. To suppress the interference, the user will send an IN request to all the interfering SBSs within distance $\mu Z_2$, which is called the \textit{IN range}, where $ \mu \in [0,+\infty)$ is referred to as the \textit{IN coefficient}, and $Z_2$ is the distance between $u_0$ and its serving SBS. 
	
	As shown in Fig.~\ref{fig: MU system Model}, all the SBSs (except the serving SBS) in the red dash line circle will receive an IN request from the typical user. When an SBS receives IN requests, it will use its available antennas to suppress its interference at the requesting users. However, due to the limited spatial DoF, each SBS can only satisfy at most $M_2 - U_2$ IN requests. Let $\varTheta_x$ be the number of IN requests received by an SBS located at $x$. If $ \varTheta_x \leq M_2-U_2 $, all the IN requests received can be satisfied; otherwise the SBS will uniformly randomly choose $M_2-U_2$ users to suppress the interference. 
	
	With this IN scheme, when $0 \leq \mu <1$ (resp. $\mu \geq 1$), all the interfering SBSs of the typical user can be divided into three parts: the interfering SBSs within the circle of radius $\mu Z_2$ (resp. $Z_2$); the interfering SBSs within the annulus from radius $\mu Z_2$ (resp. $Z_2$) to $Z_2$ (resp. $\mu Z_2$); and the remaining interfering SBSs outside the circle of radius $Z_2$ (resp. $\mu Z_2$), which correspond to Type-A, Type-B, and Type-C shown in Fig.~\ref{fig: MU system Model}, respectively. Let $\Phi_2^a$, $\Phi_2^b$, and $\Phi_2^c$ denote the set of interfering SBSs of Type-A, Type-B, and Type-C, respectively. We have $\Phi_{2}^{i} \triangleq \{ x | x \in \Phi_{2} \backslash \{ x_{20}\} , \| x \|\in \varOmega_{i} \}$, $i\in\{a,b,c\}$,
	where $\varOmega_a = [0,  \min \{Z_2, \mu Z_2 \})$, $\varOmega_b = [ \min \{ Z_2, \mu Z_2 \}, \max \{Z_2, \mu Z_2 \})$, and $\varOmega_c = [ \max \{Z_2, \mu Z_2 \}, +\infty)$ are three distance intervals.
	
	From the system model illustrated above, we can observe that the choice of $\mathcal{N}_c$, the design of $\mathbf{T}$, and the value of $\mu$ will jointly affect the performance of the network. Therefore, we will set $(\mathcal{N}_c,\mathbf{T},\mu) $ to be the design parameters.
	
	\subsection{Signal Model}\label{subsec: MU Signal Model}
	In this paper, we adopt ZFBF to serve multiple users over one time-frequency RB as well as to satisfy the IN requests for SBS-users. Perfect CSI at the transmitter is assumed. For each BS, equal power is allocated to its associated users. For the typical user $u_0$ located at the origin and served by BS at $x$, the received signal is
	\begin{align}\label{equ: MU received signal}
		y_{0 x} \! =& \sqrt{ \frac{P_ { \nu (x) } } {U_ { \nu (x) } } }  Z_{{ \nu (x) } } ^{- \frac{\alpha_{ \nu (x) } }{2}} \mathbf{h}_{0 x}^{H} \mathbf{F}_{x} \mathbf{s}_{x} \! \notag
		 \\ &+  \sum_{ x_m \! \in \tilde{\Phi}_{ \nu (x)} } \!\! \! \sqrt{ \! \frac{P_ { \! \nu ( x ) } } {U_ {\! \nu ( x  ) } } } \|x_m\|^{- \!\frac{\alpha_ { \nu (x) } }{2}} \mathbf{h}_{0 x_m}^{H} \!\! \!\mathbf{F}_{\!x_m} \mathbf{s}_{x_m} \!\!\!+\! n_0,
	\end{align}	
	where $\nu(x)$ returns the index of tier to which a BS located at $x$ belongs, i.e., $\nu(x) = k$ iff $x \in \Phi_k$; $Z_{1}$ (or $Z_{2}$) denotes the distance between $u_0$ and its serving MBS (or SBS); $\alpha_k >2$ denotes the path-loss exponent; $\tilde{\Phi}_{ \nu (x)}$ is the set of interfering BSs defined by $\tilde{\Phi}_1 = \Phi_1 \backslash \{x\}$ if $\nu (x) = 1$, and $\tilde{\Phi}_2 = \Phi_2^a \cup \Phi_2^b \cup \Phi_2^c $ if $\nu (x) = 2$;	  $\mathbf{h}_{ix}$ is the channel coefficient vector from the BS located at $x$ to the user $i$, and satisfies $\mathbf{h}_{ix} \in \mathcal{CN} (\mathbf{0}_{M_{ \nu (x) } \times 1} , \mathbf{I}_{M_ { \nu (x) } } )$; $\mathbf{s}_{x} =[s_{1x},s_{2x},\cdots, s_{ U_{ \nu (x) } x}]^{T} \in \mathbb{C}^{U_ { \nu (x) } \times 1} $ is the complex symbol vector sent by the BS located at $x$ to its $U_{ \nu (x) }$ users; $\mathbf{F}_{x} = [ \mathbf{f}_{1x},\mathbf{f}_{2x},\cdots, \mathbf{f}_{U_{ \nu (x) } x} ] \in \mathbb{C}^{M_{ \nu (x) } \times U_{ \nu (x) } } $ is the precoding matrix of BS $x$ for its $U_{ \nu (x) }$ served users, $n_0 \overset{d}{\sim} \mathcal{CN} (0, \sigma ^2) $ is the additive white Gaussian noise. 
	
	By utilizing ZFBF, each MBS can simultaneously serve $U_1$ users over one time-frequency RB; while the SBS located at $x$ can simultaneously serve $U_2$ users as well as suppress the interference at other $\min \{ \varTheta_{x} , M_2-U_2 \}$ users. The precoding vector is given by
	\begin{equation}
		\mathbf{f}_{i x}=\frac{\left(\mathbf{I}_{M_{ \nu (x) }}-\mathbf{H}_{-i x} \mathbf{H}_{-i x}^{\dagger}\right) \mathbf{h}_{i x}} {\left\| \left(\mathbf{I}_{M_{ \nu (x) }}-\mathbf{H}_{-i x} \mathbf{H}_{-i x}^{\dagger}\right) \mathbf{h}_{i x} \right\|_{2}},
	\end{equation}
	where $\mathbf{H}_{-i x} = [ \mathbf{h}_{1x}, \cdots,\mathbf{h}_{(i-1)x}, \mathbf{h}_{(i+1)x}, \cdots, \mathbf{h}_{\Lambda x}  ]$, with $\Lambda = U_{ \nu (x) } +\min\{\varTheta_{x} , M_{ \nu (x) }-U_{ \nu (x) } \} $. If the serving BS of a user located at $x$ is an MBS, $\varTheta_{x} = 0$, since there is no IN request sent to the MBS.

	Suppose $u_0$ requests file $n$ at the current time instant. Then, the received SINR at $u_0$ is
	\begin{equation}\label{equ: MU SINR}
		\Upsilon_{n,\nu (x)}=\frac{ \frac{ P_{\nu (x)} }{U_{\nu (x)} } g_{0 \nu (x) } Z_ {\nu (x)} ^{ -\alpha_{\nu (x)} }}{ \sum_{x_m \in \tilde{\Phi}_{ \nu (x)} }   \frac{P_ { \nu (x) } } {U_ { \nu (x) } }  g_{x_{\nu (x) }} \|x_m\|^{-\alpha_  { \nu (x) } }  +\sigma^{2}},
	\end{equation}
	where $x$ represents $x_{10}$ (resp. $x_{20}$) if $u_0$ is an MBS-user (resp. SBS-user); $g_{0k} \triangleq \left| \mathbf{h}_{0 x_{k0} }^H \mathbf{f}_{0 x_{k0}} \right|^2 $, $k \in \{1,2\}$, is the effective channel gain of the desired signal from $x_{k0}$, which follows $g_{0k} \overset{d}{\sim} \Gamma(D_k, 1)$ \cite{Gupta2014}, with $D_k = \max \{ {M_k - U_k + 1-\varTheta_{x_{k0}}} , 1\}$, and $\varTheta_{x_{10}} = 0;$\footnote{When $k = 1$, each MBS serves $U_1 = M_1$ users over the same time-frequency RB. Therefore, $D_1 = 1$, $g_{01} \overset{d}{\sim} \Gamma(1, 1)$, which is the exponential distribution with unit mean.} $g_{x_k} \triangleq  \left| \mathbf{h}_{0 x_m }^H \mathbf{F}_{x_m} \right|^2 $ is the interfering channel gain between $u_0$ and the BS $x_m$ from the $k$-th tier, which follows $ g_{x_k} \overset{d}{\sim} \Gamma (U_k,1)$ \cite{Gupta2014}.

	\subsection{Performance Metric}
	We use the ASE as a metric to measure the network capacity. The ASE describes the average achieved data rate per unit area normalized by the transmission bandwidth, with a unit bit/s/Hz/$\text{m}^2$, which is defined as \cite{Li2016}
	\begin{align}\label{equ: MU ASE Define}
		\mathrm{ASE} (\mathcal{N}_c,\mathbf{T},\mu) = \log _{2}(1+ \tau ) \Big( &\lambda_{1} U_{1} q_1(\mathcal{N}_c) \notag
		\\ & + \lambda_{2} U_{2} q_2 (\mathcal{N}_c, \mathbf{T}, \mu) \Big) ,
	\end{align}			
	where $\tau$ is some predefined SINR threshold, and
	\begin{equation}
		q_1(\mathcal{N}_c) \!\triangleq\! \left( \sum_{n \in \mathcal{N}_1}  a_n \! + \! \sum_{n \in \mathcal{N}_b} a_n \xi (N_b) \right)  \mathbb{P} [ \Upsilon_{n,1} \geq \tau ],
	\end{equation}
	\begin{equation}
		q_2 (\mathcal{N}_c,\mathbf{T},\mu) \triangleq  \sum_{n \in \mathcal{N}_c} a_n  \mathbb{P} \left[ \Upsilon_{n,2}  \geq \tau \right],
	\end{equation}
	represent the STP of $u_0$ as an MBS-user or an SBS-user, respectively, where $\xi (x) $ is the probability that a requested file is successfully retrieved over the backhaul link given the backhaul load is $x$. Since we consider a full-loaded network, the backhaul load of each MBS should always be full, i.e., $L_b \equiv N_b$, we have $\xi (x) = \min \left\{ 1, {C_b \over x} \right\}$. Here, the STP is an intermediate metric, which represents the probability that the received SINR at the typical user exceeds a given threshold $\tau$. 
	
	A summary of the frequently mentioned symbols is provided in Table \ref{tbl: MU symbols}.

	\section{Derivation of $\mathrm{ASE} (\mathcal{N}_c,\mathbf{T},\mu)$}\label{section: MU Performance}
	In this section, we will derive the expression for ASE under the proposed IN scheme for some given $(\mathcal{N}_c,\mathbf{T},\mu) $, and use Monte Carlo simulation to verify the analytical results.	From \eqref{equ: MU ASE Define}, we see that to compute $\mathrm{ASE} (\mathcal{N}_c, \mathbf{T}, \mu)$, it suffices to derive the expressions for the STPs $q_1(\mathcal{N}_c)$ and $q_2 (\mathcal{N}_c, \mathbf{T}, \mu)  $. Furthermore, in modern dense wireless networks, the strength of interference is much greater than that of background thermal noise. Therefore, it is reasonable to neglect the noise. In the sequel, we focus on the performance analysis of the interference-limited scenario, i.e., $\sigma^2=0$ in \eqref{equ: MU SINR}.
	\subsection{Derivation of MBS STP $q_1(\mathcal{N}_c)$} When the typical user is an MBS-user, the serving BS is its nearest MBS. Since the frequency bands used by MBS and SBS are orthogonal, there is no inter-tier interference, and all the interference comes from the MBSs that are farther away from the typical user than the serving MBS. By deriving the probability distribution function (PDF) of $ \Upsilon_{n,1}$ in \eqref{equ: MU SINR}, we can obtain the STP for the MBS tier.
	
	As shown in Appendix \ref{appendix: MU Proof prop MU q1}, the STP for the MBS tier is given by
	\begin{equation}\label{equ: MU q1}
		q_1(\mathcal{N}_c) \!= \!\left(\sum_{n \in \mathcal{N}_1} a_n \!+\! \sum_{n \in \mathcal{N}_b} a_n \xi (N_b) \right) \varPsi_1,
	\end{equation}
	with 
	\begin{equation}\label{equ: MU Psi1}
		\begin{aligned}
			\varPsi_1 = \frac{1}{  1 + F_1( \tau  )  } ,
		\end{aligned}		
	\end{equation}
	where $\xi (N_b) = \min \left\{ 1, {C_b \over N_b} \right\}$ and $F_j(x)$ is defined in \eqref{equ: MU Fj} with ${_2F_1}\left( a,b; c;d \right)$ denoting the Gauss hypergeometric function:
	\begin{equation}\label{equ: MU Fj}
		F_j(x) = {_2F_{1}}\left(-\frac{2}{\alpha_j}, U_j ; 1-\frac{2}{\alpha_j} ;-x \right)-1, \,\, j\in\{1,2\}.
	\end{equation}

	From \eqref{equ: MU q1}, we can observe that the impact of the caching policy on the STP for the MBS tier is mainly reflected by the backhaul load, which is only determined by the choice of $\mathcal{N}_b$ (or equivalently $\mathcal{N}_c$).
\begin{figure*}[!t]	
	\centering
	\subfloat[$\rho = 10$]{\includegraphics[scale=0.4]{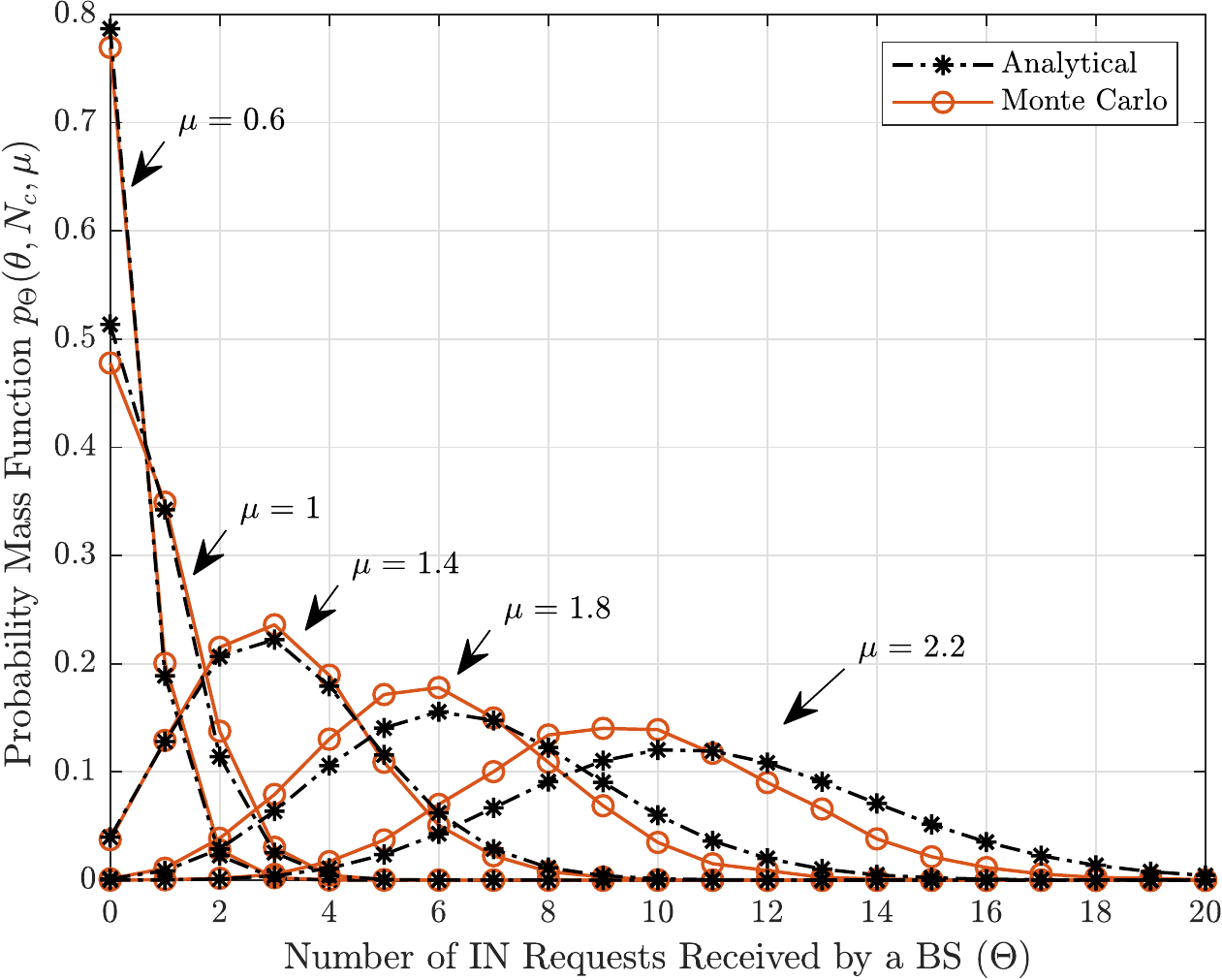} \label{fig: MU sub VerifyTheta mu} }\,\,
	\subfloat[$\mu = 1.4$] {\includegraphics[scale=0.4]{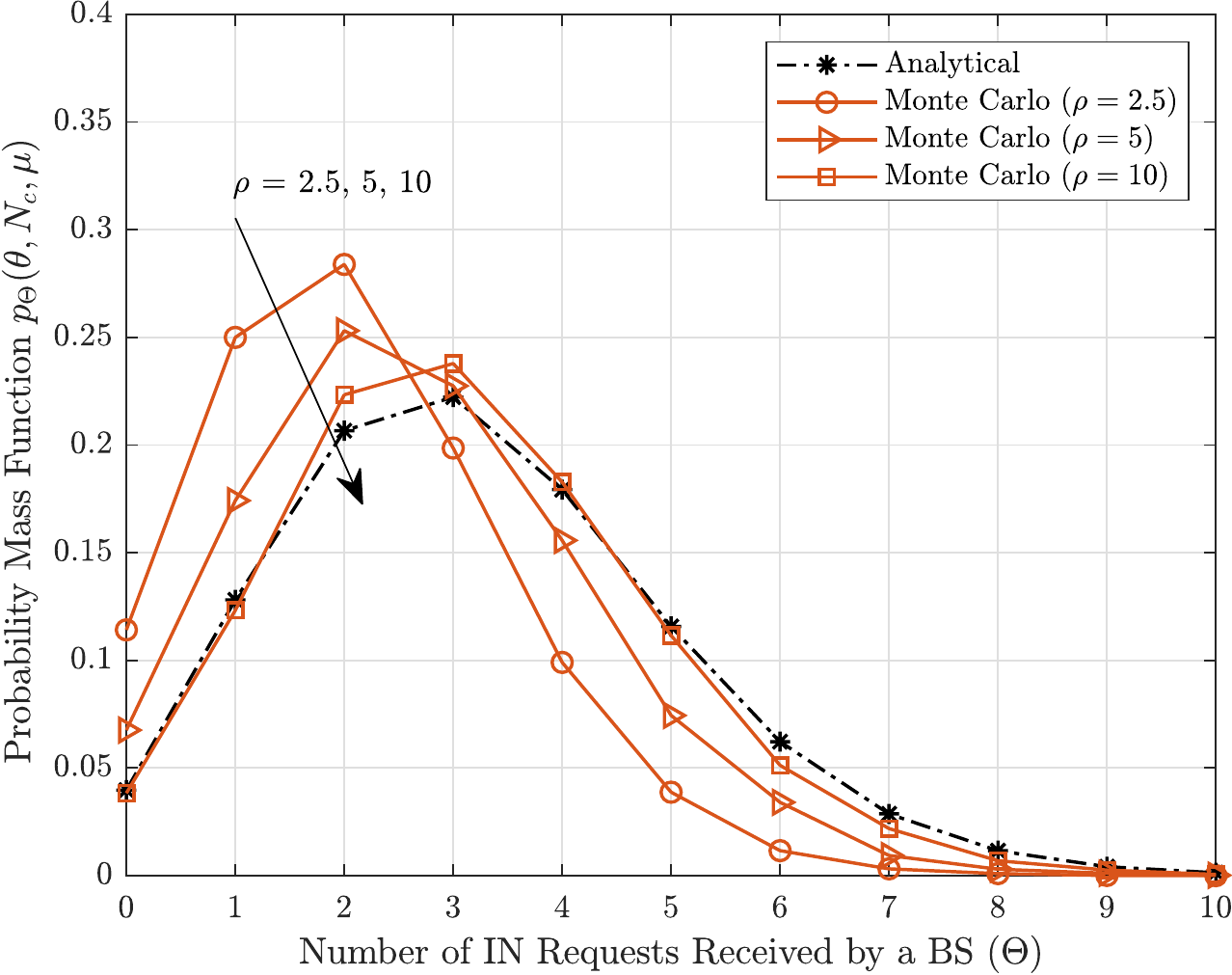} \label{fig: MU sub VerifyTheta rho}}\,\,
	\subfloat[$T_n = C_2/N_c,\, \forall n \in \mathcal{N}_c$] {\includegraphics[scale=0.4]{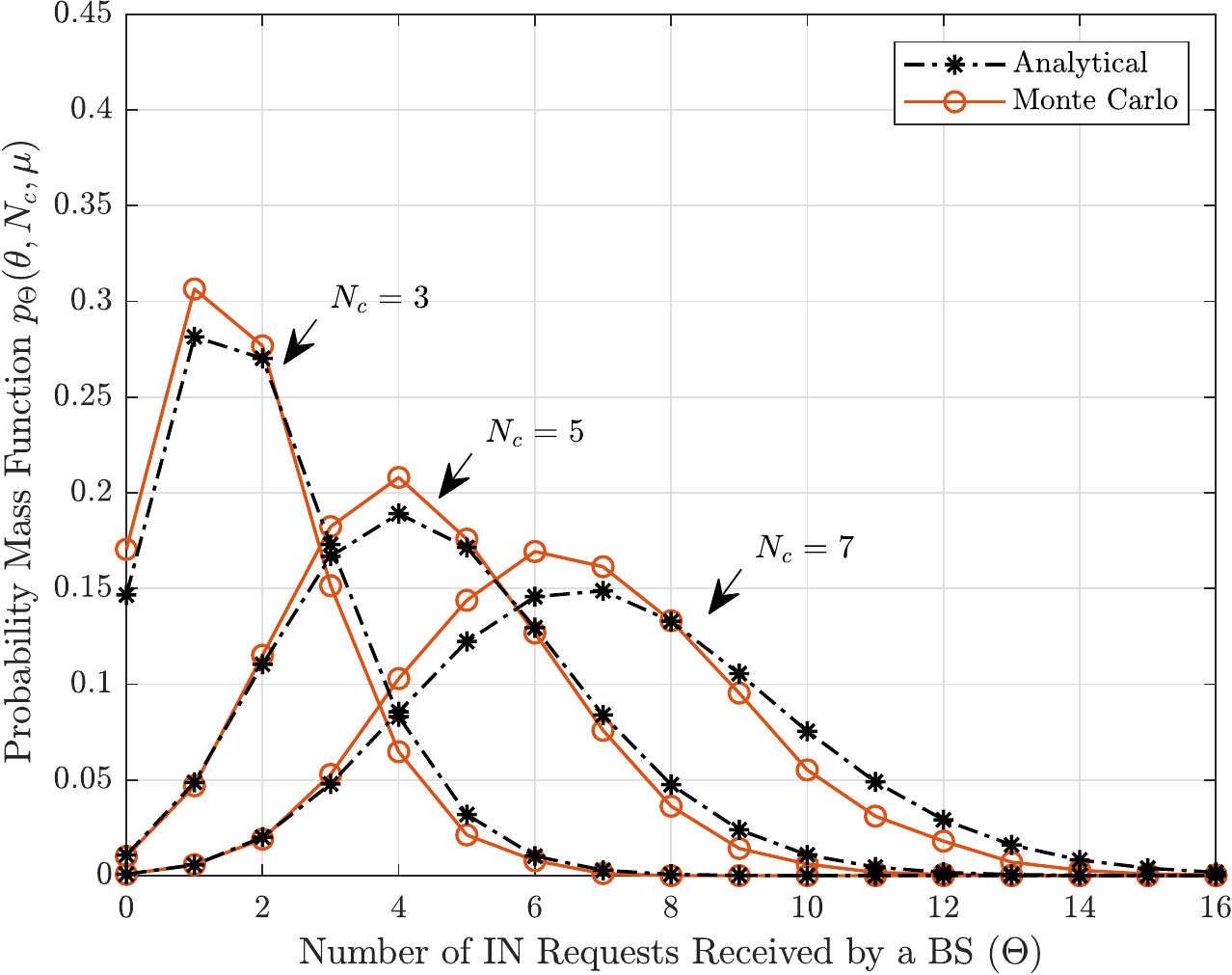} \label{fig: MU sub VerifyTheta Nc}}
	\caption{The probability mass function of $\varTheta$, i.e., $p_{\varTheta} (\theta, N_c , \mu)$. The default setting of the parameters is $M_1 = 8$, $M_2 = 6$, $U_1=8$, $U_2=2$, $P = [46,23] \, \mathrm{dBm}$, $\alpha_1 = \alpha_2 = 4$, $\lambda_{1} = 1 \times 10^{-4} \, \mathrm{m}^{-2} $, $\lambda_{2} = 5 \times 10^{-4}\, \mathrm{m}^{-2} $, $\lambda_{u} =  0.01 \, \mathrm{m}^{-2} $ (i.e., $\rho = 10$), $N=12$, $C_1 = 4$, $C_2 = 3$, $C_b = 2$, $ \mathcal{N}_1 = \{1,2,3,4 \}$, $ \mathcal{N}_c = \{5,6,7,8 \}$, $ \mathcal{N}_b = \{9,10,11,12 \}$, $\mathbf{T} = [ 0.9,0.8,0.7, 0.6 ] $, $\gamma_z = 0.8$, $\mu = 1.4$. For the sub-figure (c), uniform caching is adopted and the most popular $N_c$ files in $\mathcal{N}_2$ are cached with probability $T_n = C_2/N_c$. The Monte Carlo results are obtained by averaging $5\times 10^4$ independent random realizations with map size $2\times 2\, \mathrm{km}^2$. }
	\label{fig: MU VerifyTheta}
\end{figure*}		
	
	\subsection{Number of IN Requests Received by an SBS}
	Before deriving the SBS STP $q_2 (\mathcal{N}_c,\mathbf{T},\mu)  $, we first need to calculate the probability mass function (PMF) of the number of IN requests received by the serving SBS of a user. The content-centric association mechanism adopted in this work allows the IN coefficient $\mu$ to be smaller than 1, and the interfering SBSs are further categorized into Type-A, Type-B, and Type-C, making the network much more complicated and more challenging to analyze than that in \cite{Li2015a}.
	
	To allow feasible performance analysis of the network, we assume that 1) the users served by the SBS tier form a homogeneous PPP	$\Phi_{u}^2$, and is a thinning of $\Phi_{u}$; 2) $\Phi_{u}^2$ is independent from all the SBSs $\Phi_{2}$; and 3) the numbers of IN requests received by different SBSs are independent. Note that the first and second assumptions have been considered in previous works \cite{bai2013asymptotic,Li2015a,Cui2016a}. The third assumption is also adopted in \cite{Li2015a,Cui2016a}, where the accuracy of these three assumptions is verified. With these assumptions, the distribution of served users and that of SBSs are decoupled so that the system model becomes tractable. Our numerical results will further demonstrate that these assumptions allow accurate analysis in our system.
	
	With the above assumptions, it is clear that the number of IN requests received by an SBS, denoted by $\varTheta$, is a Poisson random variable with PMF
	\begin{equation}\label{equ: MU PMF theta}
		p_{\varTheta} (\theta, N_c , \mu) = \frac{ ( \bar{\varTheta} )^\theta }{ \theta !} e^{-\bar{\varTheta}},
	\end{equation}
	where $\bar{\varTheta}$ is the mean of the number of IN requests received by an SBS. We further observe that $\bar{\varTheta}$ is approximately given by
	\begin{equation}\label{equ: MU meanTheta}
		\bar{\varTheta} = \frac{N_cU_2\mu^2}{C_2} - \min \{ \mu^2, 1 \}U_2.	
	\end{equation}
	The detailed derivation can be found in Appendix \ref{appendix: MU proof lemma PMF Theta}.
	
	From \eqref{equ: MU PMF theta} and \eqref{equ: MU meanTheta}, we can find that the average number of requests received by an SBS is only related to the number of files cached at the SBS tier $N_c$, the cache size at each SBS $C_2$, the spatial DoF allocated for SDMA by each SBS $U_2$, and the IN coefficient $\mu$. Specifically, when $N_c$ increases, more files can be cached at SBSs, but, due to the limitation of cache size $C_2$, the probability of each file being cached will decrease, leading to a larger distance between the user and its serving SBS. Therefore, more IN requests will be received at SBSs, i.e., $\bar{\varTheta}$ increases. Similarly, increasing the cache size $C_2$ will increase the probability of files being cached, and hence, shorten the distance between users and files, so that $\bar{ \varTheta}$ decreases. Increasing $U_2$ will make more users connected to the network, resulting in a higher $\bar{ \varTheta}$. Moreover, a larger $\mu$ will make users send IN requests to more SBSs, contributing to a larger number of requests received at each SBS. In the sequel, with a slight abuse of notation, we use $p_{\varTheta} (\theta)$ to denote $p_{\varTheta} (\theta, N_c , \mu)$.
	
	The PMF in \eqref{equ: MU PMF theta} is verified by simulation shown in Fig.~\ref{fig: MU VerifyTheta}. In Fig.~\ref{fig: MU VerifyTheta}, we define $\rho \triangleq { \lambda_{u} \over \lambda_{2} U_2 }$ to characterize the user load of each SBS. From Fig.~\ref{fig: MU VerifyTheta}, we can observe that the analytical expression given in \eqref{equ: MU PMF theta} matches well with simulation especially when the load is high, and it is more accurate with smaller $\mu$ and $N_c$.

	\subsection{Derivation of SBS STP $q_2 (\mathcal{N}_c,\mathbf{T},\mu)$} 
	To obtain an expression for $q_2 (\mathcal{N}_c,\mathbf{T},\mu)$, we start from calculating the probability that an SBS has received an IN request from $u_0$ but is unable to satisfy it, which is denoted by $\varepsilon$ and referred to as the \textit{IN missing probability}. We consider an interfering SBS $B_0$ located in the IN range of $u_0$. Then it has received the request from $u_0$. Let's suppose it receives $\varTheta_0 = \theta_0$ more IN requests from other users, if $\theta_0 + 1 > M_2-U_2 $, the SBS will randomly choose $M_2 - U_2$ requests to satisfy, and hence, the request from $u_0$ will be denied with probability $ \theta_0 + 1 - (M_2 - U_2) \over \theta_0 + 1 $. We note that all the served SBS-users form a PPP and whether or not a served SBS-user sends IN request to its interfering SBSs is independent of others. Thus, given that $B_0$ has received the request from $u_0$, $\varTheta_0$ follows the same PMF as \eqref{equ: MU PMF theta}, due to Slivnyak’s theorem \cite{Haenggi2012}.
	Therefore, $\varepsilon$ is given by
	\begin{equation}\label{equ: MU epsilon}
		\varepsilon=\sum_{ \theta_0 = M_2-U_2 }^{\infty} \frac{ \theta_0 + 1 -( M_2 - U_2)} {\theta_0 + 1} p_{\varTheta}(\theta_0).
	\end{equation}		

	Then, as shown in Appendix \ref{appendix: MU proof Prop q2}, the STP for the SBS tier is
	\begin{equation}\label{equ: MU q2}
		\begin{aligned}
			q_2 (\mathcal{N}_c,\mathbf{T}, \mu) = \sum_{n \in \mathcal{N}_c} a_n  \varPsi_{2}(T_n, N_c, \mu),
		\end{aligned}
	\end{equation}
	where $\varPsi_{2}(T_n, N_c, \mu)$ is given by 
	\begin{align}\label{equ: MU Psi2}
		\varPsi_{2}&(T_n,  N_c,   \mu)   \notag
		\\ = &  T_n \!\sum_{\theta = 0}^{  M_2\! - U_2\! - 1 } \!\!  p_{\varTheta} (\theta) \! \left\|  \mathbf{W}_{D_2}(T_n, N_c, \mu)  ^{-1}   \right\|_1 \!+ \! { p_{\varTheta}^s  T_n \over T_n + w_0  } ,
	\end{align}		
	with $D_2 =  M_2-U_2+1-\theta$; $p_{\varTheta} (\theta)  $ is given by \eqref{equ: MU PMF theta};  $p_{\varTheta}^s  \triangleq \frac{\gamma( M_2-U_2 , \bar{\varTheta}) }{\Gamma (M_2-U_2 )} $ with $\bar{\varTheta}$ given by \eqref{equ: MU meanTheta}; $\gamma (s,x)=\int _{0}^{x}t^{s-1}\,\mathrm {e} ^{-t}\,{\rm {d}}t$ is the lower incomplete Gamma function;  $\mathbf{W}_{D_{2}}(T_n, N_c, \mu)$ is a $D_2 \times D_2$ lower Toeplitz matrix, given by	
	\begin{equation}\label{equ: MU matrixW2}
	\mathbf{W}_{D_{2}}(T_n, N_c, \mu) \!=\! \left[ \! \begin{array}{cccc}
	T_n \! + \!w_0  & & & \\
	-w_{1} & T_n \! + \! w_0  & & \\
	\vdots & \vdots & \ddots & \\
	-w_{D_{2}-1 } & -w_{D_{2}-2 } & \cdots & T_n \!+ \! w_0 
	\end{array} \!\right],
	\end{equation}	
	the elements of which, i.e, $w_0$ and $w_m$ for $m \geq 1$, are given by 	
	\begin{equation}\label{equ: MU w0}
	w_0 \!=   \varepsilon   \left(1  \! - \! T_{n}\right)  G \left( \tau  \right)  \! + \! a  ( 1 \! - \!  \varepsilon ) \mu^2   F_2  \left(  { \tau \over \mu ^{ \alpha_2 }}  \right)  \!+ \! b  T_n F_2(\tau ) ,
	\end{equation}	
	\begin{align}\label{equ: MU wm}
		w_m & =    \frac{ 2  (U_2)_m }{ m! } \Bigg\{\! \varepsilon \left(1 \! - \! T_{n}\right) \! \frac{1}{\alpha_2} \tau^{ \frac{2}{\alpha_2} } \! B \! \left( \! m \!-\! \frac{2}{\alpha_2}, U_2 \! + \! \frac{2}{\alpha_2} \! \right) \! \! \notag
		\\& +  a  \left(1 -  \varepsilon \right)   \mu ^{2}  \tilde{F}_{m} \left(  { \tau \over \mu^{\alpha_2} }    \right) +  b T_{n}  \tilde{F}_{m}(\tau )   \Bigg\},
	\end{align}
	with $F_j(x)$ given in \eqref{equ: MU Fj}; $a$ and $b$ are given by
	\begin{equation} \label{equ: MU a and b}
		a = \left\{
		\begin{array}{ll}
		1-T_n &  \,\, \mu<1, \\
		1 & \,\, \mu\geq 1,\\
		\end{array}
		\right. \quad
		b = \left\{
		\begin{array}{ll}
		1 &  \,\, \mu<1, \\
		\varepsilon & \,\, \mu\geq 1;\\
		\end{array}
		\right.
	\end{equation}
	$G(x)$ is given by	
	\begin{equation}\label{equ: MU Gj}
		G(x) = \frac{ \Gamma \left(1-\frac{2} {\alpha_2 } \right) \Gamma \left(U_{2}+ \frac{2} {\alpha_2} \right) } {\Gamma \left( U_{2} \right)}  x  ^{ \frac{2} {\alpha_2}};
	\end{equation}
	$(U)_m = U(U+1)\cdots(U+m-1)$ is the Pochhammer symbol; $B(x,y)$ is the Beta function; and $\tilde{F}_{k}(x)$ is
	\begin{equation}\label{equ: MU Fjk}			
		\tilde{F}_{k}(x) \!=\!  \frac{x^k}{\alpha_2 k - 2}  {_2F_{1}} \left( \! U_2 \! + \! k, k \! - \! \frac{2}{\alpha_2};  k \!- \! \frac{2}{\alpha_2} \!+ \! 1 ; \!-x \! \right).
	\end{equation}
	
	From the expression for $q_2 (\mathcal{N}_c,\mathbf{T},\mu)$, we can observe that the STP of the SBS tier is affected by the design parameters $(\mathcal{N}_c , \mathbf{T}, \mu)$ in an extremely complicated manner. Specifically, the number of elements in $\mathcal{N}_c$, i.e., $N_c$, and the IN coefficient $\mu$ will affect $\bar{ \varTheta}$, so that the PMF $p_{\varTheta} (\theta)$ and the IN missing probability $\varepsilon$ are affected. Serving as the weights of summation in \eqref{equ: MU Psi2}, $p_{\varTheta} (\theta)$ will exert an influence on $\varPsi_{2}(T_n,N_c, \mu)$ directly. Whereas $\varepsilon$ will affect the Toeplitz matrix $\mathbf{W}_{D_2} (T_n ,N_c, \mu)$ firstly, and then puts an impact on $\varPsi_{2}(T_n,N_c, \mu)$. In terms of $T_n$, however, it only affects $\mathbf{W}_{D_2} (T_n ,N_c, \mu)$ directly and has no influence on $\bar{ \varTheta}$. Note that the result given in \eqref{equ: MU q2} is a generalized version of that in \cite{Li2015a} with further considering the random caching policy and SDMA at the SBSs. By plugging $T_n = 1$ for $\forall n \in \mathcal{N}_c$, and $U_2 = 1$ into \eqref{equ: MU q2}, the expression for $q_2 (\mathcal{N}_c,\mathbf{T},\mu)$ degrades into the STP obtained in \cite{Li2015a}. 
	
	We note that, by combining \eqref{equ: MU q1}, \eqref{equ: MU q2}, and \eqref{equ: MU ASE Define}, we can obtain the final expression for $\mathrm{ASE} (\mathcal{N}_c,\mathbf{T},\mu)$.

	\subsection{Special Case Where $U_2=M_2$} 
	In this subsection, we consider a special case where all the available DoF of each SBS is used for SDMA, so that there is no antenna allocated for IN, i.e., $U_2 = M_2$. In this case, we can obtain a much simplified closed-form expression for the ASE. In particular, the conditional STP $ \varPsi_{2}(T_n, N_c, \mu) $ degrades into 
	\begin{equation}\label{equ: MU Psi2 U2=M2}
		\varPsi_{2}^{sp}(T_n) = { T_n \over ( 1 - G(\tau) + F_2 (\tau) ) T_n + G( \tau ) }.
	\end{equation}
	By replacing $\varPsi_{2}(T_n, N_c, \mu)$ with $ \varPsi_{2}^{sp} (T_n)$ in \eqref{equ: MU q2}, and considering \eqref{equ: MU ASE Define} and \eqref{equ: MU q1}, $  \mathrm{ASE} ^{sp} (\mathcal{N}_c, \mathbf{T} )$ can be obtained.
	
	Note that \eqref{equ: MU Psi2 U2=M2} has a similar formation to Lemma 4 in \cite{Cui2017a}, where a single-input single-output network (with $M_k=U_k=1$ for all $k\in \{1,2\}$) is considered. From \eqref{equ: MU Psi2 U2=M2}, we can observe that given $\mathcal{N}_c$, the design parameter $T_n$ is separated from other network parameters such as $\alpha_2$, $U_2$, and $\tau$. Besides, $ \mathrm{ASE} ^{sp} (\mathcal{N}_c, \mathbf{T} ) $ is a concave function with respect to (w.r.t.) $T_n$, for $\forall n \in \mathcal{N}_c$. Therefore, given $\mathcal{N}_c$, when optimizing $\mathbf{T}$ to maximize the metric $ \mathrm{ASE} ^{sp} (\mathcal{N}_c, \mathbf{T} ) $, the complexity of the solution can be significantly reduced by using KKT conditions. Unfortunately, optimizing the ASE for the general case of $U_2 < M_2$ is much harder, as will be shown in Section~\ref{section: MU Max}.
	\subsection{Numerical Validation} 
\begin{figure}[!t]	
		\centering
		\subfloat[SINR Threshold, $\mu = 1$]{
			\includegraphics[scale=0.31]{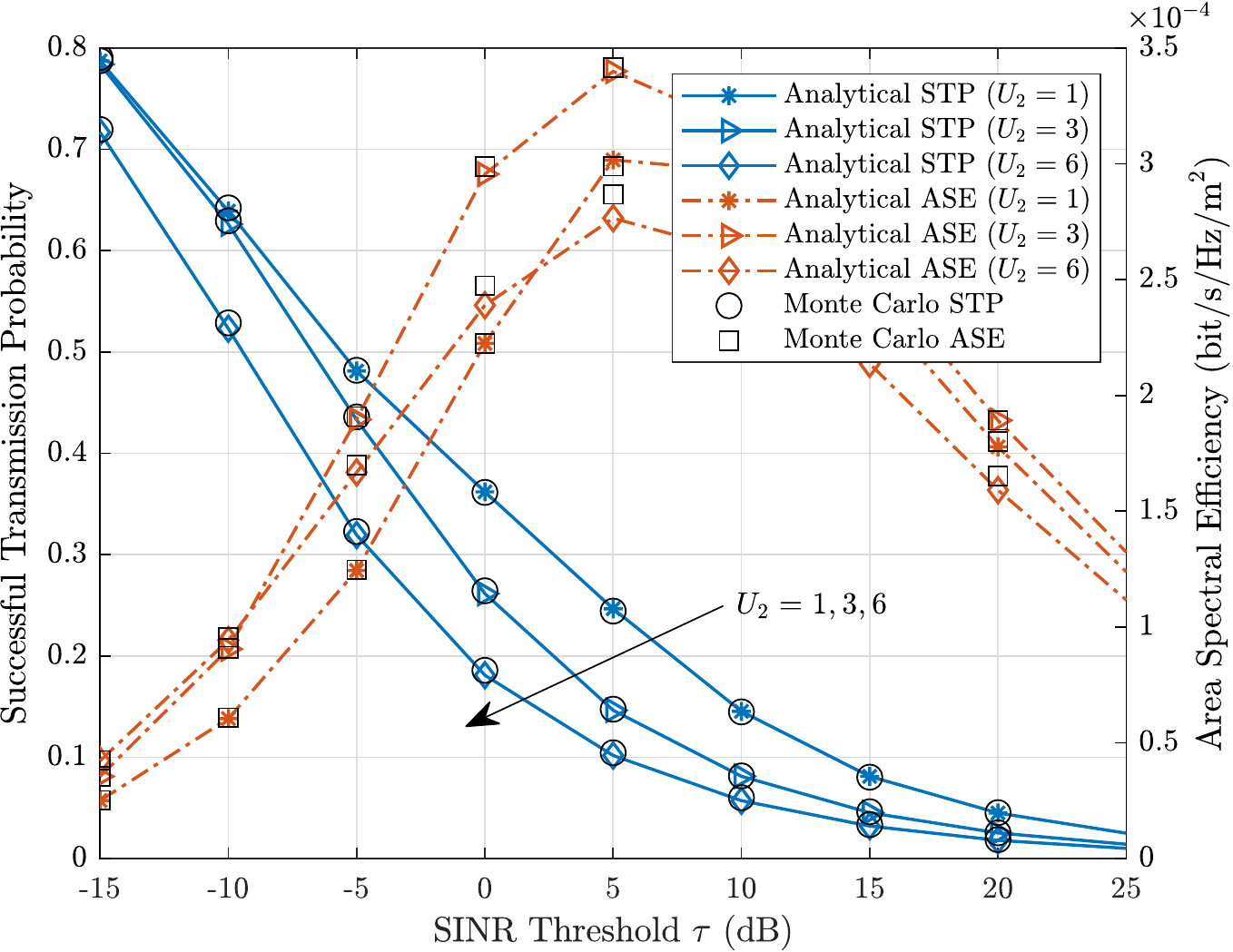}\label{fig: MU sub SA vs Tau} }
		\subfloat[IN Coefficient, $\tau = 0\,\mathrm{dB}$]{
			\includegraphics[scale=0.31]{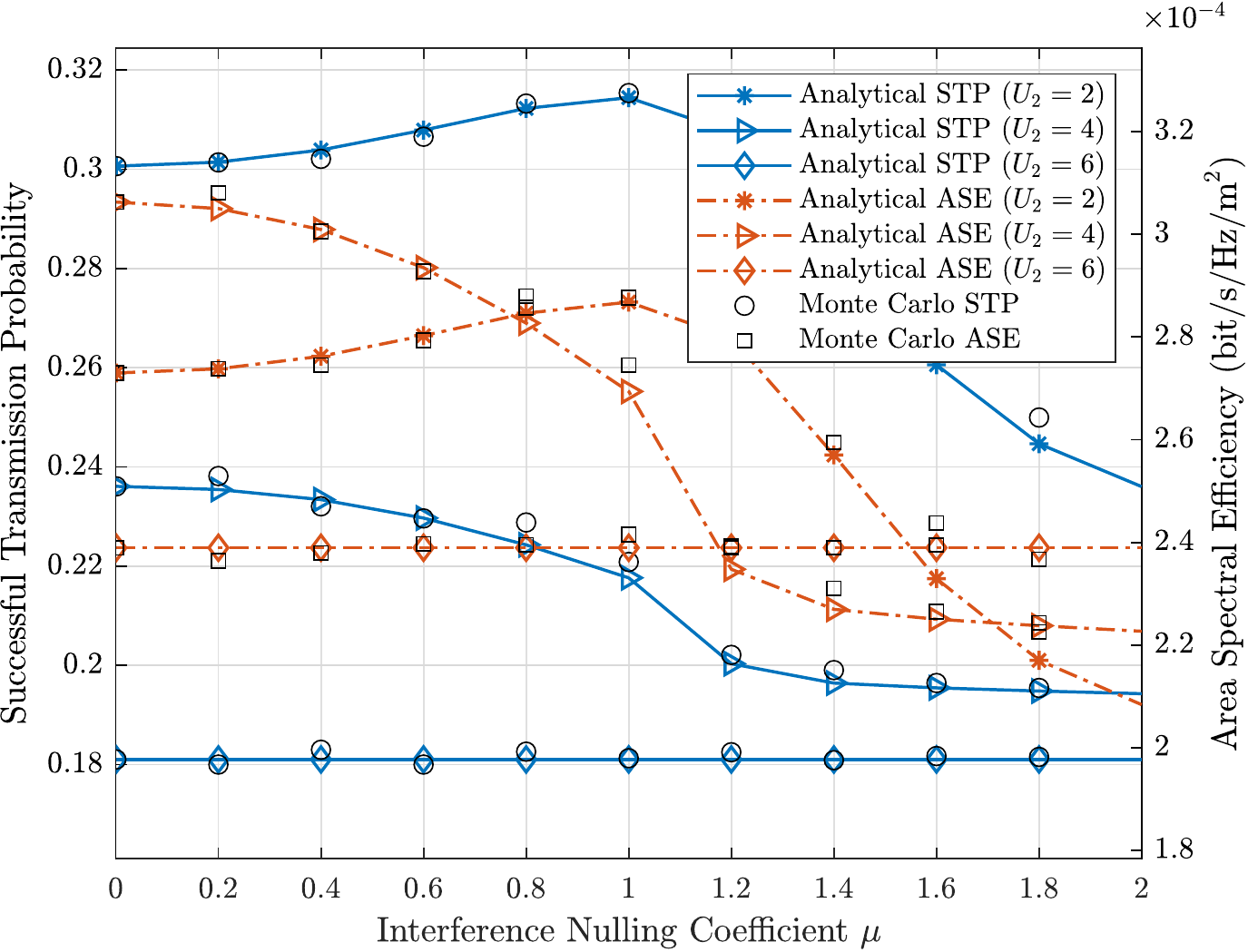}\label{fig: MU sub SA vs Mu}}
		\caption{The STP and ASE versus SINR threshold $\tau$ and IN coefficient $\mu$. The simulation parameters are set to be the same as in Fig. \ref{fig: MU VerifyTheta}.}
		\label{fig: MU VerifySTPASE_TAUMU}
	\end{figure}

	For numerical validation, Fig.~\ref{fig: MU VerifySTPASE_TAUMU} plots the total STP and ASE versus the SINR threshold $\tau$, the IN coefficient $\mu$, and the number of users $U_2$ served simultaneously by an SBS. Here, the total STP is defined as $q(\mathcal{N}_c,\mathbf{T}, \mu) = q_1 (\mathcal{N}_c) + q_2 (\mathcal{N}_c,\mathbf{T}, \mu)$. We compare the above ASE analysis with Monte Carlo simulation results.
	Fig.~\ref{fig: MU VerifySTPASE_TAUMU} shows that the analytical expressions for $q  (\mathcal{N}_c, \mathbf{T}, \mu)$ and $ \mathrm{ASE} (\mathcal{N}_c,\mathbf{T},\mu)$ match the simulation results, even though some approximations have been used (cf. Section \ref{section: MU Performance}). Therefore, in the sequel, we will focus on the analysis and optimization for the analytical expressions obtained in this section.
	Fig.~\ref{fig: MU VerifySTPASE_TAUMU}\subref{fig: MU sub SA vs Mu} shows that for the special case where $U_2=M_2$, the STP and ASE are independent of $\mu$; otherwise, the variation tendency of the STP with $\mu$ is similar to that of the ASE with $\mu$, and $\mu$ should be carefully designed to achieve optimal network performance.
	\section{Area Spectral Efficiency Maximization}\label{section: MU Max}
	In this section, we will maximize the ASE by jointly optimizing the parameters $(\mathcal{N}_c, \mathbf{T}, \mu)$. Specifically, the optimization problem can be formulated as follows.
	\begin{problem}[ASE Optimization Problem] \label{Prob: MU Original Problem}
		\begin{subequations}
			\begin{align}
				\mathrm{ASE} ^\star  & = \underset{\mathcal{N}_c,\mathbf{T},\mu}{\max} \,\, \mathrm{ASE} (\mathcal{N}_c,\mathbf{T},\mu )  \notag \\
				\text{s.t.}
				& \quad  N_c \geq C_2, \label{equ: MU Prob Con1}\\
				& \quad  0\leq T_n \leq 1 ,   \,\,\, \sum_{n \in \mathcal{N}_c} \! T_{n} = C_2,\label{equ: MU Prob Con2}\\
				& \quad \mu \geq 0. \label{equ: MU Prob Con3}
			\end{align}		
		\end{subequations}
	\end{problem}

	In the sequel, we use $\mathcal{N}_c^\star$, $\mathbf{T}^\star$, and $\mu^\star$ to denote the optimal value of variables $\mathcal{N}_c$, $\mathbf{T}$, and $\mu$, respectively. By optimizing the cache placement parameters $\mathcal{N}_c$ and $\mathbf{T}$ as well as the IN coefficient $\mu$, the objective function in Problem \ref{Prob: MU Original Problem} can be maximized. However, the feasible sets for $ \mathbf{T} $ and $\mu$ are continuous whereas that for $\mathcal{N}_c$ is discrete, making Problem \ref{Prob: MU Original Problem} a mixed integer programming problem. To effectively solve Problem~\ref{Prob: MU Original Problem}, we next partition it into two sub-problems, i.e., \textit{cache placement optimization problem} and \textit{IN coefficient optimization problem}, which will be solved alternately.		
	\subsection{Cache Placement Optimization}\label{subsec: MU Caching Optimization}
	First, we consider the cache placement optimization problem with a given $\mu$, which can be formulated as follows.
	\begin{problem}[Cache Placement Optimization]\label{Prob: MU Cache Placement}
		\begin{align}
		\mathrm{ASE}  ^\star (\mu)  & = \underset{\mathcal{N}_c,\mathbf{T}}{\max} \,\, \mathrm{ASE} (\mathcal{N}_c,\mathbf{T},\mu )   \notag \\
		\text{s.t.}
		& \quad  \eqref{equ: MU Prob Con1}, \eqref{equ: MU Prob Con2}.
		\end{align}			
	\end{problem}		
	Note that Problem \ref{Prob: MU Cache Placement} is still a mixed integer programming problem where the discrete part is to choose the elements of the set $\mathcal{N}_c$, and the continuous part is to design the caching probability vector $\mathbf{T}$. 
	
	We will decompose the ASE into the MBS-tier and SBS-tier components as $\mathrm{ASE}_1 = \lambda_{1} U_{1} q_1 (\mathcal{N}_c)  \log _{2}(1+ \tau ) $ and $\mathrm{ASE}_2 =  \lambda_{2} U_{2} q_2 (\mathcal{N}_c,\mathbf{T}, \mu) \log _{2}(1+ \tau )$. Then, Problem \ref{Prob: MU Cache Placement} has the following equivalent form:
	\begin{problem}[Equivalent Problem]\label{Prob: MU Equivalent Problem}
		\begin{align}
		\mathrm{ASE}^{\star}(\mu) &= \underset{\mathcal{N}_c} {\max} \,\,\mathrm{ASE}_1 (\mathcal{N}_c) + \mathrm{ASE}_{2}^{\star}(\mathcal{N}_c, \mu )  \notag \\
		\text{s.t.}
		& \quad  \eqref{equ: MU Prob Con1}, \label{equ: MU Prob Equivalent 1}
		\end{align}		
		\begin{align}
		\setlength\abovedisplayskip{0pt}
		\mathrm{ASE}_{2}^{\star}(\mathcal{N}_c, \mu) &= \underset{\mathbf{T} } {\max} \,\, \mathrm{ASE}_{2} (\mathcal{N}_c,\mathbf{T},\mu ) \notag \\
		\text{s.t.}
		& \quad \eqref{equ: MU Prob Con2}. \label{equ: MU Prob Equivalent 2}
		\end{align}		
	\end{problem}

	For the discrete optimization problem in \eqref{equ: MU Prob Equivalent 1}, there are totally $\sum_{N_c=C_2}^{N_2}{\binom{N_2}{N_c}}=O(N_2^{N_2})$ different choices in the feasible set. Even if $\mathrm{ASE} _{2}^{\star} (\mathcal{N}_c , \mu )$ is given for each choice of $\mathcal{N}_c$, the computation complexity can be prohibitive when $N_2$ and $N_c$ are large. On the other hand, for the continuous optimization problem, it is hard to determine the convexity of the objective function in \eqref{equ: MU Prob Equivalent 2}. In the sequel, we will separately investigate the discrete part and the continuous part of Problem \ref{Prob: MU Equivalent Problem}, and explore some properties of the ASE to simplify the optimization problem.
	
	\subsubsection{Discrete Optimization}
	The aim of the discrete optimization is to determine the optimal file set $\mathcal{N}_c^\star$. To reduce the total number of choices that need to be considered, we observe the following property.
	\begin{property}\label{proper: MU Nb consecutive}
		In the interference-limited scenario, when the network is full-loaded, there exists an optimal	$\mathcal{N}_b^\star$ in which the indexes of files are consecutive, i.e., $\mathcal{N}_b^\star = \{ n_1^b, n_1^b+1, \cdots, n_1^b + N_b^\star \}$, with $N_b^\star = | \mathcal{N}_b^\star |$, and $n_1^b \in \{ N_1+1, \cdots, N - N_b^\star \} $ being the index of the first file in $\mathcal{N}_b^\star$.
	\end{property} 
	\begin{IEEEproof}
		Please refer to Appendix~\ref{appendix: MU proof proper Nb consec}.
	\end{IEEEproof}

	Property \ref{proper: MU Nb consecutive} can significantly reduce the difficulty of solving the discrete optimization. By applying this property, noting that $\mathcal{N}_c = \mathcal{N}_2 \backslash \mathcal{N}_b$, the total number of choices that need to be considered for $\mathcal{N}_c$ is cut down to $\sum_{N_b=1}^{N_2-C_2}{\sum_{n_{1}^{b}=N_1+1}^{N-C_2-N_b+1}{1}=O\left( N_{2}^{2} \right)} $. This allows us to use exhaustive search to solve the discrete optimization problem without losing optimality.

	\subsubsection{Continuous Optimization}
	{\color{black}The objective function of the continuous problem shown in \eqref{equ: MU Prob Equivalent 2} is complicated, whose convexity is hard to determine. However, $\mathrm{ASE}_{2} (\mathcal{N}_c,\mathbf{T},\mu )$ is a differentiable function w.r.t. $\mathbf{T}$, and the constraint set is convex. Therefore, we can use the gradient projection method (GPM) \cite{DimitriP.Bertsekas2016} to obtain a stationary point of the continuous problem. Even though GPM can solve the continuous problem properly, the convergence rate is highly dependent on the choice of the stepsize, which, however, may degrade the efficiency of the algorithm when selected improperly. Moreover, when calculating the derivative, an inverse of the Toeplitz matrix needs to be calculated, which will inevitably increase the computational complexity.
		
	Therefore, to reduce the computation complexity in solving the continuous problem, we may consider using the following simple lower bound $ \mathrm{ASE}^l (\mathcal{N}_c,\mathbf{T},\mu) $ as the objective function for Problem \ref{Prob: MU Original Problem}. 
	
	As shown in Appendix \ref{appendix: MU proof theorem Two Bounds}, a lower bound on $\varPsi_{2}(T_n, N_c, \mu)$ is given by
	\begin{align}
	\varPsi_{2}^l  (T_n, N_c, \mu) 
	= \!\!  \sum_{\theta = 0}^{  M_2-U_2 - 1 }     { p_{\varTheta} (\theta)T_n \over {\varrho}_1  T_n + {\varrho}_2 } +    { p_{\varTheta}^s   T_n \over \varrho_{1,1} ( 1 ) T_n +   \varrho_{2,1} ( 1 ) } , \label{equ: MU Psi2 lower}
	\end{align}
	where $\beta = (D_2 !)^{-\frac{1}{ D_2 }}$, $D_2 =  M_2-U_2+1-\theta $, and $\varrho_{1} $, $\varrho_{2} $, $\varrho_{1,i}(x) $, and $\varrho_{2,i}(x) $ are given by
	\begin{align}
	{\varrho}_{1} & = \left\{
	\begin{aligned}
	&\begin{aligned}
	\varrho_{1,1}(1) +  &\sum_{m = 1}^{D_2-1} \frac{ 2  (U_2)_m }{ m! }  \left( 1 - { m \over D_2} \right) \\ & \times \left( \tilde{\varrho}_{m}(\tau) - \tilde{F}_{m}(\tau ) \right)
	\end{aligned}
	&   \mu<1, \\
	&\begin{aligned}
	\varrho_{1,1}(1)   +   \varepsilon  &\sum_{m = 1}^{D_2-1} \frac{ 2  (U_2)_m }{ m! } 
	\left( 1  -  { m \over D_2} \right) \\& \times \left( \tilde{B}_m(\tau)  -   \tilde{F}_{m}(\tau )  \right) 
	\end{aligned}
	&  \mu\geq 1,				 
	\end{aligned}\right. \label{equ: MU varrho1} \\
	{\varrho}_{2} & =  \varrho_{2,1}(1) - \sum_{m = 1}^{D_2-1} \frac{ 2  (U_2)_m }{ m! } \left( 1  -  { m \over D_2} \right) \tilde{\varrho}_{m} (\tau) , 			\label{equ: MU varrho2}
	\end{align}
	\begin{align}
	\varrho_{1,i}( x ) &= \left\{
	\begin{aligned}
	&1 - \varrho_{2,i}( x ) +  F_2(  i x \tau ) &  \,\, \mu<1, \\
	&1 - \varepsilon G \left(  i x \tau  \right)  + \varepsilon F_2(  i x \tau ) & \,\, \mu\geq 1,
	\end{aligned}
	\right.		\label{equ: MU varrho1ix}\\
	\varrho_{2,i}( x ) &= \varepsilon G \left(  i x \tau  \right) + (1 - \varepsilon) \mu^2   F_2\left({  i x \tau \over \mu ^{ \alpha_2 }} \right),  \label{equ: MU varrho2ix}		
	\end{align}	
	with $\tilde{B}_m(\tau) = \frac{1}{\alpha_2} \tau^{ \frac{2}{\alpha_2} }  B (  m - \frac{2} {\alpha_2}, U_2 + \frac{2}{\alpha_2} )$, and $\tilde{\varrho}_{m}(\tau) =\varepsilon 	\tilde{B}_m(\tau)  +  (1  -   \varepsilon )  \mu ^{2 }  \tilde{F}_{m} ( { \tau \over \mu^{\alpha_2} } )$.
	
	By replacing the $\varPsi_{2}(T_n, N_c, \mu)$ with $\varPsi_{2} ^l (T_n, N_c, \mu)$ in \eqref{equ: MU q2}, a lower bound on the ASE, i.e., $ \mathrm{ASE}^l (\mathcal{N}_c,\mathbf{T},\mu)$ can be obtained. Compared with the expression for $\mathrm{ASE} (\mathcal{N}_c,\mathbf{T},\mu)$ given in Section \ref{section: MU Performance}, when we calculate this bound, the calculation of the inverse of the Toeplitz matrix $ \mathbf{W}_{D_{2}}(T_n, N_c, \mu) $ is avoided, thereby reducing the computational burden. Moreover, $\varrho_{1}$, $\varrho_{2}$, $\varrho_{1,i}(x) $, and $\varrho_{2,i}(x) $ are functions of variables $( N_c, \mu)$, but are independent of $T_n$. This will ease the design of $\mathbf{T}$ when $\mathcal{N}_c$ is given.
	
	Note that Property~\ref{proper: MU Nb consecutive} still holds here, which can be proved by defining $f(x,y) \triangleq \min \{ 1, {C_b \over N_2 - y} \}  \lambda_{1} U_1 \varPsi_{1} - \lambda_{2} U_2 \varPsi_{2}^l( x , y, \mu )$ in Appendix~\ref{appendix: MU proof proper Nb consec} and following the similar procedure. Given $\mathcal{N}_c$ and $\mu$, the objective function in \eqref{equ: MU Prob Equivalent 2} will be replaced by its corresponding lower bound, denoted by $\mathrm{ASE}_{2}^{l} (\mathcal{N}_c, \mathbf{T}, \mu )$. Therefore, the continuous optimization problem can be rewritten as follows. 	
	\begin{problem}[Lower Bound Optimization]\label{Prob: MU Lowerbound Problem}		
		\begin{align}
		 \underset{\mathbf{T} } {\max} \,\,  & \mathrm{ASE}_2^l (\mathcal{N}_c,\mathbf{T},\mu) \notag \\
		 &\text{s.t.} \quad \eqref{equ: MU Prob Con2}.
		\end{align}		
	\end{problem}
	
	It can be easily observed that the optimization in Problem~\ref{Prob: MU Lowerbound Problem} is a convex problem. As shown in Appendix \ref{appendix: MU KKT}, by using the KKT conditions, we can obtain the optimal solution to it as	
	\begin{equation} \label{equ: MU Solution to LowerBound}
		T_{n}^{ \dagger}=\left\{\begin{array}{ll}{0,} & {a_n f(0) < \nu^\dagger  ,} \\ {1,} & { a_n f(1) >  \nu^\dagger ,} \\ {x( T_{n}^{  \dagger}, \nu^\dagger ),} & {\text { otherwise, }}\end{array}\right.
	\end{equation}
	where 
	\begin{align}\label{equ: MU derivative f}
		f(x) &\triangleq \left.{  \partial  \mathrm{ASE}^l (\mathcal{N}_c,\mathbf{T},\mu) \over a_n \partial T_n } \right|_{T_n = x} =  \lambda_{2} U_2 \log_{2} ( 1 + \tau) \notag
		\\ & \times \bigg[ \sum_{\theta = 0}^{  M_2-U_2 - 1} \!\!\!  {   p_{\varTheta} (\theta ) \varrho_{2} \over  ( \varrho_{1}  x \! + \!  \varrho_{2}  )^2 } \! + \!  { p_{\varTheta}^s  \varrho_{2,1} ( 1 )  \over ( \varrho_{1,1} ( 1 ) x \! + \! \varrho_{2,1} ( 1 ) )^2 }  \bigg], 
	\end{align}
	where $x( T_{n}^{ \dagger}, \nu^\dagger )$ denotes the root of equation $a_n f( x ) = \nu^\dagger  $, and the optimal Lagrangian multiplier $\nu^\dagger $ satisfies $\sum_{n \in \mathcal{N}_c} T_{n}^{ \dagger} ( \nu^\dagger ) = C_2$. In addition, both $x( T_{n}^{ \dagger}, \nu^\dagger )$ and $\nu^\dagger $ can be found by a simple bisection search.
	\subsection{IN Coefficient Optimization}\label{subsec: MU IN Optimization}
	In this part, we consider the following IN coefficient optimization problem given the solution of Problem \ref{Prob: MU Cache Placement} is obtained.
	\begin{problem}[IN Coefficient Optimization]\label{Prob: MU IN coefficient opt}
		\begin{align}
		\mathrm{ASE} ^\star  & = \underset{ \mu }{\max} \,\,\mathrm{ASE} ^\star (\mu)  \notag \\
		\text{s.t.}
		& \quad  \eqref{equ: MU Prob Con3},
		\end{align}			
	\end{problem}	
	where $\mathrm{ASE}^\star (\mu) $ is obtained by solving Problem \ref{Prob: MU Cache Placement}.

	Problem \ref{Prob: MU IN coefficient opt} is a one-dimensional optimization problem with only an orthant constraint \cite{DimitriP.Bertsekas2016}. We can effectively obtain its optimal solution using line search. Furthermore, in practice, the number of available antennas for IN at each SBS is limited. Therefore, the value of $\mu$ should be upper-bounded to prevent the value of $ \varTheta $ from being too large so that the IN scheme can perform effectively. To this end, we impose a constraint on $\mu$ to satisfy $ \bar{ \varTheta} \leq M_2-U_2 $. Note that a similar constraint is considered in \cite{Hosseini2018}. By substituting \eqref{equ: MU meanTheta} into $ \bar{ \varTheta} \leq M_2-U_2 $, we obtain the following constraint:		 	
	\begin{equation}\label{equ: MU all IN condition}
		\left\{\begin{aligned}					
		&0 \leq \mu \leq \sqrt{ \delta_A \over \delta_F}, & \text{if} \,\,\delta_A \geq \delta_F, \\
		&0 \leq \mu \leq \sqrt{ \delta_A -1 \over \delta_F - 1 }, & \text{if} \,\, \delta_A < \delta_F,			
		\end{aligned}
		\right.
	\end{equation}
	where $\delta_A \triangleq M_2 / U_2 \geq 1$ and $\delta_F \triangleq N_c / C_2 \geq 1$ represent the antenna gain and file diversity gain, respectively. It should be noted that \eqref{equ: MU all IN condition} is not a necessary constraint for solving Problem \ref{Prob: MU IN coefficient opt}. However, with this constraint, the process of solving Problem \ref{Prob: MU IN coefficient opt} will become more efficient.
	\begin{algorithm}
	\caption{Stationary Point for $ \mathrm{ASE}^l (\mathcal{N}_c,\mathbf{T},\mu) $ Maximization}
	\label{Algo: MU Original Problem}
	\begin{algorithmic}[1]	
		\STATE Initialize $i=0$, $\mu^{(0)\dagger} \in [0, \sqrt{\delta_A}] $, and $\mathrm{ASE} ^\dagger = 0$.
		\REPEAT
		\STATE Fix $\mu^{(i)\dagger}$.
		\FOR{$N_c \in \{C_2, C_2+1, \cdots, N_2\} $}
		\FOR{all $ \mathcal{N}_c$ satisfying $|\mathcal{N}_c | = N_c$ and Property \ref{proper: MU Nb consecutive}}
	\STATE Obtain $T_{n}^{ \dagger}$ for $\forall n\in \mathcal{N}_c$ according to \eqref{equ: MU Solution to LowerBound}. 
		\STATE $\mathbf{T}^{\dagger}(\mathcal{N}_c) \leftarrow [T_{n}^{ \dagger}]_{n \in \mathcal{N}_c} $, compute $\mathrm{ASE} ^ \prime \triangleq \mathrm{ASE}^l ( \mathcal{N}_c, \mathbf{T}^{\dagger} ( \mathcal{N}_c),\mu^{(i)\dagger})$. 
		\IF{ $\mathrm{ASE} ^\dagger \leq \mathrm{ASE} ^ \prime $} \label{step: MU 1}
		\STATE Set $ ( \mathcal{N}_c^{(i+1)\dagger}, \mathbf{T}^{(i+1)\dagger}, \mathrm{ASE} ^\dagger ) =  (\mathcal{N}_c,  \mathbf{T}^{\dagger} ( \mathcal{N}_c) , \mathrm{ASE} ^ \prime)$.
		\ENDIF			\label{step: MU 2}
		\ENDFOR
		\ENDFOR	
		\STATE Fix $ (\mathcal{N}_c^{(i+1)\dagger}, \mathbf{T}^{(i+1)\dagger} )$, obtain the optimal solution $ \mu^{\dagger} $ to Problem \ref{Prob: MU IN coefficient opt} with line search, and compute $\mathrm{ASE} ^ \prime \triangleq \mathrm{ASE}^l(\mathcal{N}_c^{(i+1)\dagger}, \mathbf{T}^{(i+1)\dagger}, \mu^{\dagger})$.
		\IF{$\mathrm{ASE} ^\dagger \leq \mathrm{ASE} ^ \prime $}\label{step: MU 3}
		\STATE {Set $(\mu^{(i+1)\dagger}, \mathrm{ASE} ^\dagger) = (\mu^{ \dagger}, \mathrm{ASE}  ^ \prime )$.}
		\ENDIF \label{step: MU 4}
		\STATE $i \leftarrow i+1$.
		\UNTIL Convergence.
	\end{algorithmic}
\end{algorithm}
	\subsection{Alternating Optimization}\label{subsection: MU Alternating Opt}
	Algorithm~\ref{Algo: MU Original Problem} summarizes the whole procedure of alternately solving the cache placement optimization problem in Section~\ref{subsec: MU Caching Optimization} and the IN coefficient optimization problem in Section~\ref{subsec: MU IN Optimization}. In each iteration, by using the KKT conditions, an optimal solution of Problem \ref{Prob: MU Lowerbound Problem} can be obtained; by using exhaustive search, we can obtain an optimal solution for the discrete problem in \eqref{equ: MU Prob Equivalent 1}. Therefore, when considering $\mathrm{ASE}^l (\mathcal{N}_c,\mathbf{T},\mu) $ as the objective function, we can obtain an optimal solution $(\mathcal{N}_c ^\dagger,\mathbf{T}^\dagger)$ of Problem~\ref{Prob: MU Cache Placement}. In addition, given $(\mathcal{N}_c ^\dagger,\mathbf{T} ^\dagger)$, an optimal solution $\mu^\dagger$ for Problem~\ref{Prob: MU IN coefficient opt} can be obtained using line search. The Step~\ref{step: MU 1} $\sim$ Step~\ref{step: MU 2} and Step~\ref{step: MU 3} $\sim$ Step~\ref{step: MU 4} in Algorithm \ref{Algo: MU Original Problem} are used to ensure a solution not worse than the current one in each iteration when alternately solving Problem~\ref{Prob: MU Cache Placement} and Problem~\ref{Prob: MU IN coefficient opt}. Therefore, when we set $\mathrm{ASE}^l (\mathcal{N}_c,\mathbf{T},\mu) $ as the objective function in Problem~\ref{Prob: MU Original Problem}, the alternating procedure in Algorithm \ref{Algo: MU Original Problem} converges to a stationary point of this problem \cite[pp. 268]{DimitriP.Bertsekas2016}. We denote it by $(\mathcal{N}_c ^\dagger,\mathbf{T}^\dagger,\mu^\dagger)$. We take this stationary point as our \textit{proposed policy} for maximizing $\mathrm{ASE} (\mathcal{N}_c,\mathbf{T},\mu )$ in Problem~\ref{Prob: MU Original Problem}.

\section{Performance Evaluation}\label{section: MU Numerical}	
	In this section, we evaluate the proposed joint hybrid caching policy and user-centric IN scheme in Matlab simulation.
	\subsection{Comparison Benchmarks}
	Note that Fig.~\ref{fig: MU VerifySTPASE_TAUMU}\subref{fig: MU sub SA vs Mu} in Section~\ref{section: MU Performance} provides a comparison of the system performance with (i.e., $\mu \neq 0$) and without IN (i.e., $\mu = 0$), indicates the advantages of our work with IN scheme over the works in \cite{Liu2017a, Kuang2019, Kuang2019a}, where interference management is not considered. In addition, \cite{Jiang2019c} and \cite{Xu2019c} consider single-tier networks, which is different from our two-tier network layout. Hence, it is difficult to draw an equivalent performance comparison between our work and those two works due to the incompatibility of network settings. Therefore, we omit further performance comparison between our work and \cite{Liu2017a, Kuang2019, Kuang2019a,Jiang2019c,Xu2019c }.
	
	Instead, we compare the proposed method with two benchmarks that employ both caching and IN, MPC \cite{Bastug2015} and UDC \cite{Tamoor-ul-Hassan2015a}. For both benchmarks, each MBS selects the $ C_1$ most popular files from the whole content library to store. The remaining files, forming the sub-set $\mathcal{N}_2$, need to be further separated into two sets, whether to be served by SBSs or by MBSs via backhaul links. In MPC, each SBS selects the $C_2$ most popular files from $\mathcal{N}_2$ to store, and the remaining files in $\mathcal{N}_2$ are served by MBSs via backhaul links. In UDC, the $C_b$ least popular files of $\mathcal{N}_2$ are served by MBSs via backhaul links, and each SBS randomly selects $C_2$ different files from the remaining $N_2 - C_b$ files to store, according to the uniform distribution. For each of the benchmarks, the IN scheme is adopted, where the optimal IN coefficient is obtained by solving Problem \ref{Prob: MU IN coefficient opt} with the corresponding given caching policy $(\mathcal{N}_c,\mathbf{T})$. 
	
	Moreover, for performance benchmarking, we further consider an upper bound of $ \mathrm{ASE} (\mathcal{N}_c, \mathbf{T}, \mu)$, denoted by $ \mathrm{ASE}^u (\mathcal{N}_c, \mathbf{T}, \mu)$, as shown in Appendix~\ref{appendix: MU CCP}. By replacing the objective function of Problem~\ref{Prob: MU Original Problem} with $ \mathrm{ASE}^u (\mathcal{N}_c, \mathbf{T}, \mu)$, we obtain an upper bound optimization problem, a stationary point of which, denoted by $(\mathcal{N}_c^\ddagger, \mathbf{T}^\ddagger, \mu^\ddagger)$, can be reached efficiently using the convex-concave procedure (CCP) approach \cite{Lipp2016}, as illustrated in Appendix~\ref{appendix: MU CCP}. We repeat the procedure 5 times with different random initial values and choose the maximum $ \mathrm{ASE}^u (\mathcal{N}_c^\ddagger, \mathbf{T}^\ddagger, \mu^\ddagger)$ to obtain an approximate solution to the global optimum w.r.t. the upper bound. The result is shown with the legend ``Upper Bound'' in Figs.~\ref{fig: MU sub ASE-tau Baseline}-\ref{fig: MU sub ASE-U2 Baseline}. Thus, even though we cannot obtain an optimal solution to Problem~\ref{Prob: MU Original Problem}, we can evaluate how close the performance of our proposed policy is to the optimal one due to the relation $ \mathrm{ASE} (\mathcal{N}_c^\dagger, \mathbf{T}^\dagger, \mu^\dagger) \leq \mathrm{ASE}^\star (\mathcal{N}_c^\star, \mathbf{T}^\star, \mu^\star) \leq \mathrm{ASE}^u (\mathcal{N}_c^\ddagger, \mathbf{T}^\ddagger, \mu^\ddagger)$.

	\subsection{Simulation Results}
	Unless otherwise stated, the simulation settings are as follows: $M_1 = 32$, $M_2 = 16$, $U_1=32$, $U_2=4$, $P_1 = 46 \, \mathrm{dBm}$, $P_2 = 23 \, \mathrm{dBm}$, $\alpha_1 = \alpha_2 = 4$, $\lambda_{1} = 1 \times 10^{-4} \, \mathrm{m}^{-2} $, $\lambda_{2} = 5 \times 10^{-4}\, \mathrm{m}^{-2} $, $\lambda_{u} =  0.01 \, \mathrm{m}^{-2} $, $\tau = 0 \, \mathrm{dB}$, $N=50$, $C_1 = 20$, $C_2 = 10$, $C_b = 3$, $\gamma_z= 0.4$. Figs.~\ref{fig: MU sub ASE-tau Baseline}-\ref{fig: MU sub ASE-U2 Baseline} plot the ASE versus the SINR threshold $\tau$, the Zipf exponent $\gamma_z$, the backhaul capacity $C_b$, and the number of users served $U_2$. From Figs.~\ref{fig: MU sub ASE-tau Baseline}-\ref{fig: MU sub ASE-U2 Baseline}, we find that the maximum relative performance gap between our proposed policy (i.e., $ \mathrm{ASE} (\mathcal{N}_c^\dagger, \mathbf{T}^\dagger, \mu^\dagger)$) and the upper bound (i.e., $\mathrm{ASE}^u (\mathcal{N}_c^\ddagger, \mathbf{T}^\ddagger, \mu^\ddagger)$) is smaller than $8\%$, which indicates that the performance of our proposed policy is close to $\mathrm{ASE}^\star (\mathcal{N}_c^\star, \mathbf{T}^\star, \mu^\star)$. Furthermore, the proposed caching policy outperforms MPC and UDC. 
	
	More specifically, from Fig.~\ref{fig: MU sub ASE-tau Baseline}, we see that when $\tau$ is small (resp. large), the performance of the proposed method is almost the same as that of UDC (resp. MPC). This is because when $\tau$ is small, the SINR threshold is easy to reach, so file transmission can be successful even if the serving SBS of the typical user is not its nearest one, and hence caching more files in the network can increase the probability that the requested files are found in the SBSs. When $\tau$ is large, the typical user can only be served successfully by its nearest SBS, so ensuring the most popular files are successfully served is a better choice, which results in the least number of files cached in the SBSs.
	\begin{figure}[!t]	
		\centering
		\includegraphics[scale=0.4]{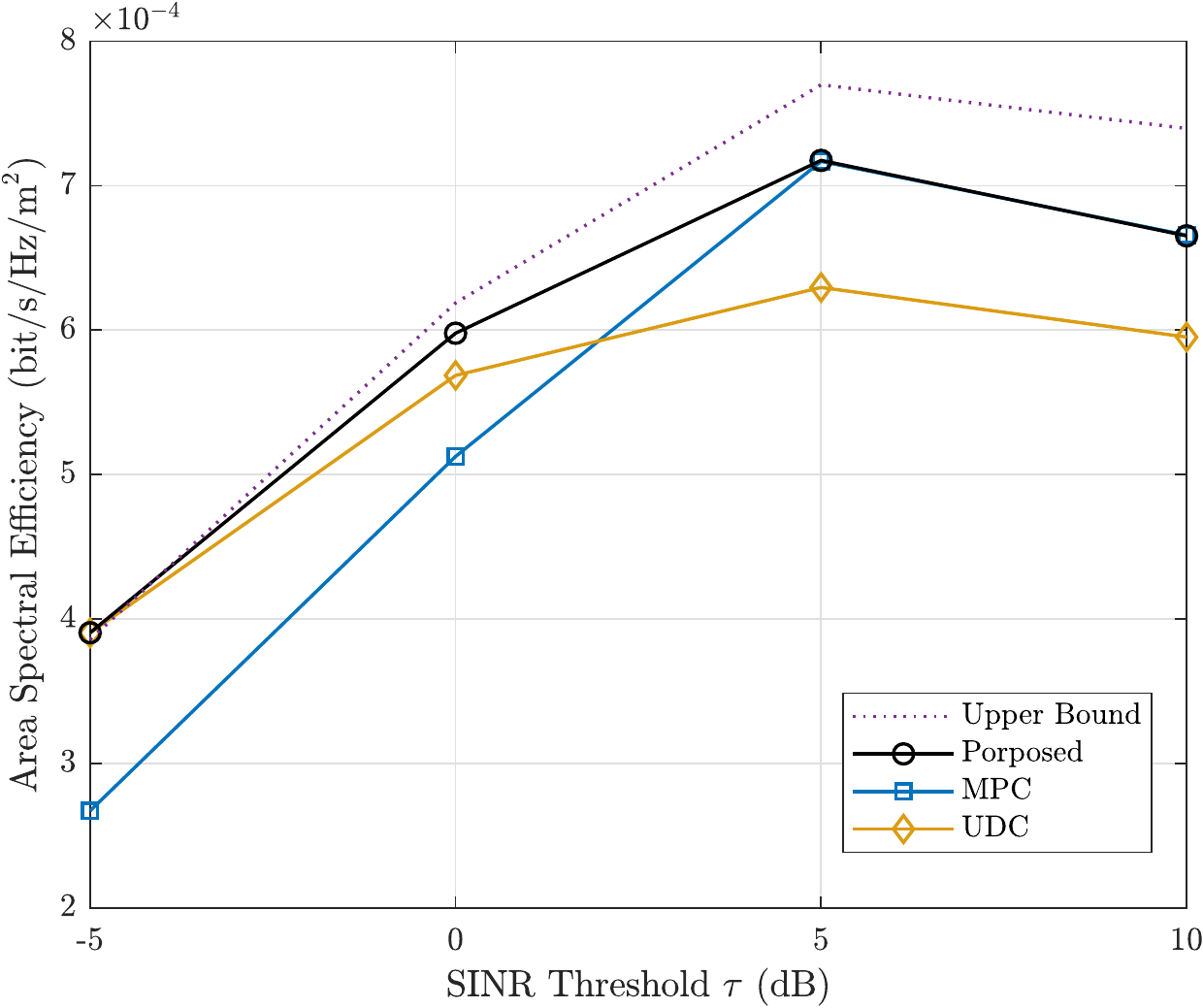}
		\caption{ASE versus SINR threshold $\tau$.}
		\label{fig: MU sub ASE-tau Baseline}
	\end{figure}
	\begin{figure}[!t]	
		\centering
		\includegraphics[scale=0.4]{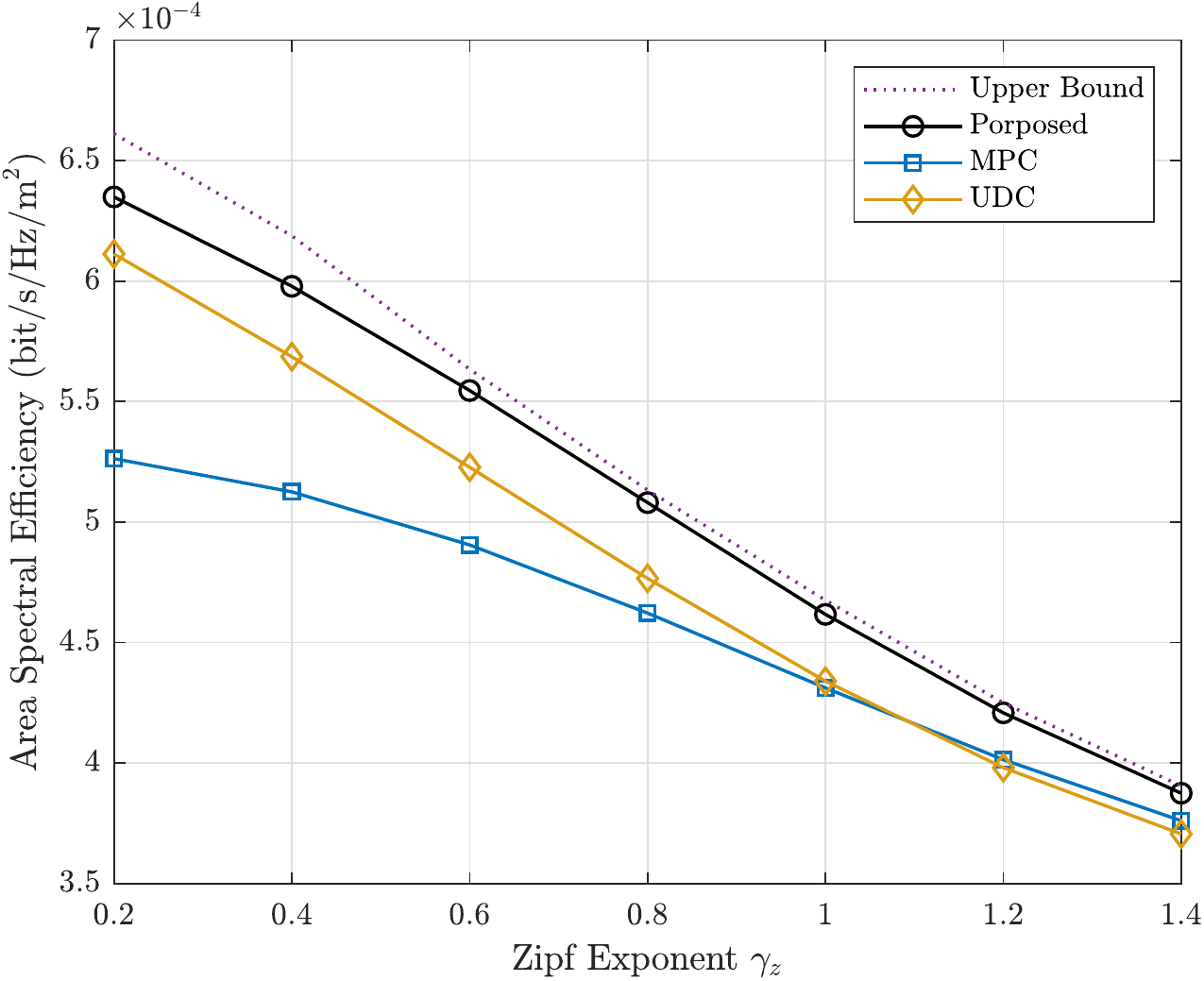}
		\caption{ASE versus Zipf exponent $\gamma_z$.}
		\label{fig: MU sub ASE-gamma Baseline}
	\end{figure}
	\begin{figure}[!t]	
		\centering
		\includegraphics[scale=0.4]{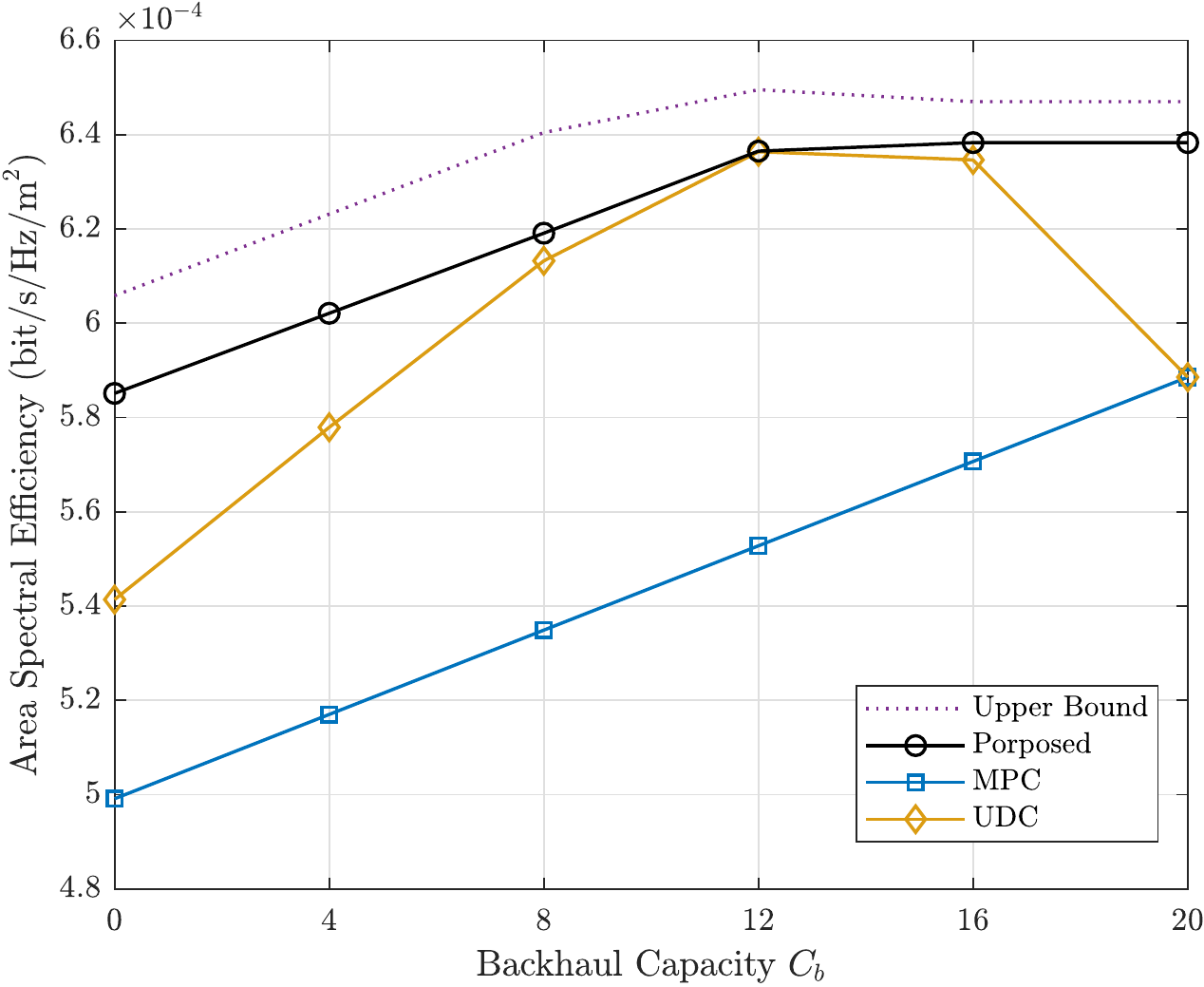}
		\caption{ASE versus backhaul capacity $C_b$.}
		\label{fig: MU sub ASE-Cb Baseline}
	\end{figure}
	\begin{figure}[!t]	
		\centering
		\includegraphics[scale=0.4]{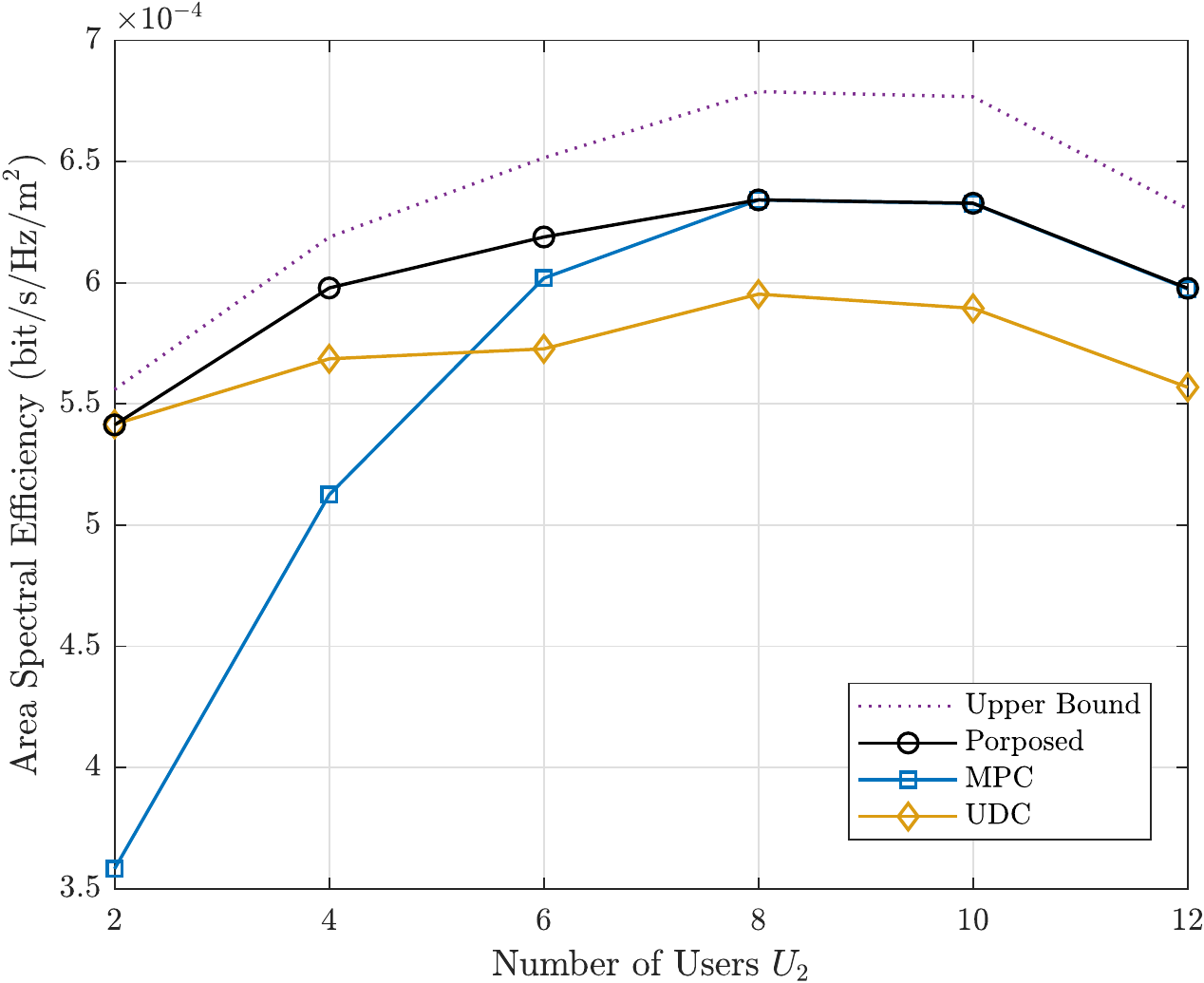}
		\caption{ASE versus number of users served $U_2$.}
		\label{fig: MU sub ASE-U2 Baseline}
	\end{figure}

	From Fig.~\ref{fig: MU sub ASE-gamma Baseline}, we observe that the ASE of the network decreases with $\gamma_z$. This is different from the results in \cite{Cui2017a,Kuang2019,Zhang2020}, where the network performance increases with $\gamma_z$. As a matter of fact, due to the dense deployment of the SBSs and the IN scheme adopted in our work, the ASE of the MBS tier is lower than that of the SBS tier. When $\gamma_z$ increases, the requests of users are more concentrated on the most popular files, which, according to our hybrid caching policy, are mainly stored at MBSs. Hence, more users are served by the MBS tier, leading to a decrease in ASE. Therefore, the proposed method gives higher ASE when the file popularity distribution is flatter.
	
	Fig.~\ref{fig: MU sub ASE-Cb Baseline} indicates that an increase of $C_b$ generally leads to higher ASE for the proposed method and MPC, since the larger $C_b$ is, the more files can be retrieved via backhaul links. When $C_b=40$, the UDC policy lets each SBS choose $C_2 = 20$ files from $\mathcal{N}_2$ (of which the number of files is $20$) to store, which is exactly the same as what the MPC policy does. However, the ASE of the proposed policy in this case is higher than that of UDC and MPC, since in the proposed policy, more files can be chosen to be served by SBSs rather than by backhaul links. In this case, the backhaul resource will not be fully utilized.
	
	Fig.~\ref{fig: MU sub ASE-U2 Baseline} depicts the relationship between the ASE and $U_2$. We observe that when $U_2$ is small, the performance of UDC is close to that of the proposed method, whereas, with $U_2$ increasing, the performance of MPC gets closer to that of the proposed method. The reason for this is that when $U_2$ is small, almost all the IN requests received by SBSs can be satisfied, so caching more files in the SBSs leads to better performance. However, an increase of $U_2$ leads to an increase of IN missing probability $\varepsilon$, which degrades performance. To offset this negative effect, fewer files should be cached in the network, according to \eqref{equ: MU meanTheta}, which leads to $N_c = C_2$, corresponding to MPC.
\section{Conclusion}\label{section: MU Conclusion}
	In this paper, we have considered random caching and IN design for cache-enabled multi-user multi-antenna HetNets. We derive the ASE over a two-tier network with hybrid file caching at both the MBSs and SBSs, where each SBS-user also sends an IN request to the interfering SBSs within its IN range for interference suppression. We obtain a simple lower-bound expression for the ASE. Then, we optimize the caching policy and the IN coefficient toward ASE maximization. By partitioning the problem into two sub-problems and exploiting the properties of the ASE, an alternating optimization algorithm is proposed to obtain a stationary point. Our numerical results show that the proposed solution is close to the optimum, and it achieves significant performance gains over existing caching policies.
	
\appendices
\section{Derivation of $q_1(\mathcal{N}_c)$}\label{appendix: MU Proof prop MU q1}
	To derive $q_1(\mathcal{N}_c)$, we need to calculate the complementary cumulative distribution function $ \mathbb{P} \left[ \Upsilon_{n,1}  \geq \tau \right]$, where $\Upsilon_{n, 1}=  g_{0 1 } Z_ {1} ^{ -\alpha_{1} } /I_1  $ is the received signal-to-interference ratio at $u_0$, with $Z_{1}$ being the distance between $u_0$ and its serving MBS $x_{10}$ and $I_1  = \sum_{x \in \Phi_1 \backslash x_{10}  }   g_{x_{1 }} \|x\|^{-\alpha_  { 1 } }$. The derivation for $ \mathbb{P} \left[ \Upsilon_{n,1}  \geq \tau \right]$ has already been studied in \cite{Andrews2011,Haenggi2012}, we restate it here for completeness.
	Conditioning on $Z_1 = z_1$, we have
	\begin{equation}\label{equ: MU Appendix CCDF for SINR1}
	\mathbb{P} \left[ \Upsilon_{n,1}  \geq \tau \right]	\!=  \! \int_{0}^{\infty} \!\!  \mathbb{P} [  \Upsilon_{n,1} \geq \tau \mid Z_1 = z_1 ]  f_{Z_1} (z_1) \mathrm{d}z_1,
	\end{equation}
	where $f_{Z_1} (z_1)= 2\pi \lambda_1 z_1 e ^{-\pi \lambda_1 z_1^2}$ is the PDF of $Z_1$. We have
	\begin{align}\label{equ: MU Appendix Conditional CCDF for SINR1}
		& \mathbb{P} [ \left. \Upsilon_{n,1} \geq \tau \right| Z_1 =z_1 ]
		 = \mathbb{E}_{I_1 }  \left[ \mathbb{P} \left[ g_{01} \geq \tau z_1^{\alpha_1}  I_1  \right] \right] \notag
		\\& \overset{(a)}{=}  \mathbb{E}_{I_1 }   [ \exp ( - \tau z_1 ^{\alpha_1}  I_1  ) ]
		=  \mathcal{L}_{I_1} ( \tau z_1^{\alpha_1} \!),
	\end{align}
	where (a) is due to $g_{01} \overset{d}{\sim} \Gamma(D_1, 1) = \Gamma(1,1) = \text{Exp}(1) $; and $\mathcal{L}_{I_1} (\cdot)$ is the Laplace transform of $I_1$, which is given by
	\begin{align}	\label{equ: MU Appendix LI1}
		\mathcal{L}_{I_1} ( s ) &= \mathbb{E}_{\Phi_{1}, g_{x_1}} \left[ \exp\left( -s \sum_{x \in \Phi_{1}\backslash  x_{10}  } g_{x_1} \| x \| ^{-\alpha_1}  \right) \right] \notag
		\\&  \overset{(a)}{=} \mathbb{E}_{\Phi_{1}} \left[ \prod_{x \in \Phi_{1}\backslash  x_{10}  } \mathbb{E}_{ g_{x_1}} \left[ \exp \left( -s  g_{x_1} \| x \| ^{-\alpha_1}  \right) \right] \right] \notag
		\\ & \overset{(b)}{=}\mathbb{E}_{\Phi_{1}} \left[ \prod_{x \in \Phi_{1}\backslash  x_{10}  }  (1 + s \| x \|^{-\alpha_1})^{-U_1} \right] \notag
		\\& \overset{ (c) } {=} \exp  \left(    -2\pi\lambda_1   \int_{z_1} ^{\infty}  \left( 1- (1 + s v^{-\alpha_1})^{-U_1} \right) v \mathrm{d}v \right) \notag
		\\& = \exp \left( - \pi \lambda_1 z_1^2 F_1(s z_1^{-\alpha_1}) \right),
	\end{align}
	where $s =\tau z_1^{\alpha_1}$ and $F_1(x)$ is given by \eqref{equ: MU Fj}; (a) is due to the independence of different small-scale fading channels; (b) follows from $ g_{x_1} \overset{d}{\sim} \Gamma (U_1,1)$; (c) is from the probability generating functional for a PPP \cite{Haenggi2012}, and considering the conversion from Cartesian coordinate to polar coordinate. Substituting \eqref{equ: MU Appendix Conditional CCDF for SINR1} and \eqref{equ: MU Appendix LI1} into \eqref{equ: MU Appendix CCDF for SINR1}, and using $\int_{0}^{\infty} 2xe^{-Ax^2} \mathrm{d}x = 1/A $, we can obtain $\varPsi_1$ as \eqref{equ: MU Psi1}.

\section{Derivation of $\bar{\varTheta}$}\label{appendix: MU proof lemma PMF Theta}
	Denote by $\Phi_{u,n}^\prime$ the users that request file $n$ and are served by the SBS tier, the density of which is denoted by $\lambda_{u,n}^\prime$. Since $\Phi_{u}^2 $ is a PPP and the file choices are independently made, $\Phi_{u,n}^\prime$ is a thinned PPP of $\Phi_{u}^2$. Let $ \mathcal{I}_i^n$ be the $i$-th combination from $\mathcal{I}^n$ (recalling that $\mathcal{I}^n$ is the set of combinations containing file $n$). Consider a typical SBS $B_0$ located at the origin and storing file $n$. The probability that it stores $ \mathcal{I}_i^n$ is ${p_i \over T_n}$ (with $i \in \mathcal{I}^n$). Denote by $p(n| \mathcal{I}_i^n)$ the probability that a randomly chosen user associated with an SBS that stores $ \mathcal{I}_i^n$ requests file $n$. Since there are totally $U_2$ users associated with $B_0$, the average number of connected users that request file $n$ is $ \bar{U}_{2,n} \! = \!\sum_{i \in \mathcal{I}^n } {p_i \over T_n} U_2 p(n| \mathcal{I}_i^n)$. 
	\begin{figure}[!t]	
		\centering
		\subfloat[$i = 1$]{
			\includegraphics[scale=0.33]{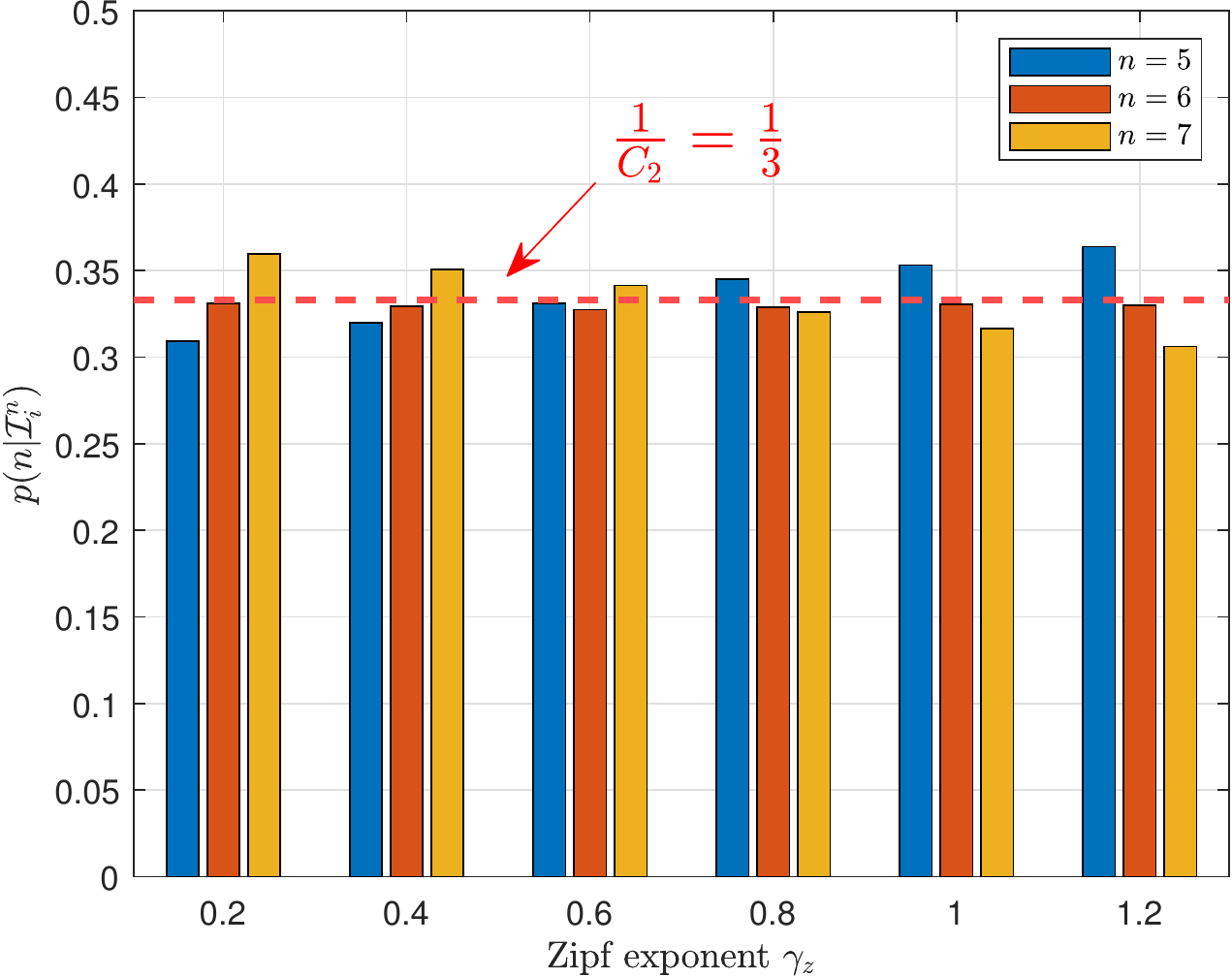}\label{fig: MU sub APB i=1} }
		\subfloat[$n=5$]{
			\includegraphics[scale=0.33]{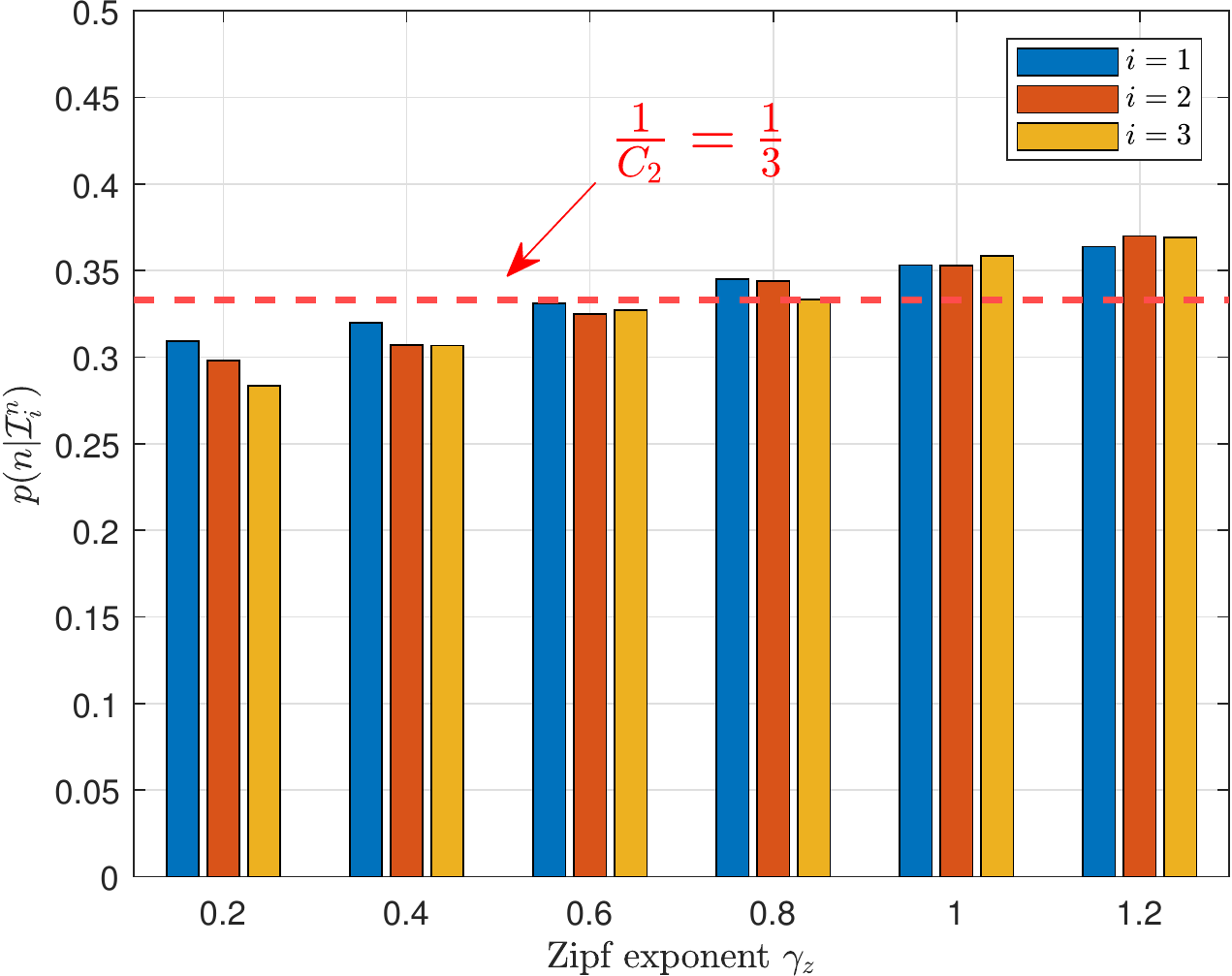}\label{fig: MU sub APB n=5}}
		\caption{Verification of the uniform distribution approximation for $p(n| \mathcal{I}_i^n)$. The simulation parameters are set to be the same as in Fig.~\ref{fig: MU VerifyTheta}. In this case, totally $I = \binom{4}{3} = 4$ file combinations are considered, i.e., $ \mathcal{I}_1 = \{5,6,7\}$, $ \mathcal{I}_2 = \{5,6,8\}$, $ \mathcal{I}_3 = \{5,7,8\}$, $ \mathcal{I}_4 = \{6,7,8\}$, whose corresponding caching probabilities 
		are $p_1 = 0.4$, $p_2 = 0.3$, $p_3 = 0.2$, $p_4 = 0.1$, respectively. For the sub-figure (a), $ \mathcal{I}_1^5 = \mathcal{I}_1^6 =\mathcal{I}_1^7 = \mathcal{I}_1 $; for the sub-figure (b), $ \mathcal{I}_1^5 = \mathcal{I}_1 $, $ \mathcal{I}_2^5 = \mathcal{I}_2 $, and $ \mathcal{I}_3^5 = \mathcal{I}_3 $.}
		\label{fig: MU VerifyAppB}
	\end{figure}	
	
	Considering the density of SBSs that store file $n$ is $\lambda_{2} T_n$, we have $\lambda_{u,n}^\prime = \lambda_{2} T_n \bar{U}_{2,n} = \lambda_{2} U_2  \sum_{i \in \mathcal{I}^n } p_i p(n| \mathcal{I}_i^n)$. We can see that $\lambda_{u,n}^\prime$ highly depends on the combinations of files containing file $n$, i.e., $\mathcal{I}^n$, which has size $I^n = \binom{ N_c -1}{ C_2-1}$. When $N_c$ and $C_2$ are large, it is unrealistic to calculate all the elements in $\mathcal{I}^n$. To this end, we use the uniform distribution to approximate $p(n| \mathcal{I}_i^n)$, i.e., $p(n| \mathcal{I}_i^n) \approx 1/C_2$. The accuracy of this approximation is verified by simulation in Fig.~\ref{fig: MU VerifyAppB}.\footnote{Given a file set $\mathcal{N}_2$ and its corresponding caching probability vector $\mathbf{T}$, we can obtain a set of caching probabilities, i.e., $p_i$, $\forall i \in \mathcal{I}$, corresponding to all the file combinations in $\mathcal{I}$, in the sense of least-squares, by solving the following optimization problem: ${\min}_{p_i, \forall i \in \mathcal{I} }  \sum_{n\in\mathcal{N}_c } (\sum_{i \in \mathcal{I}^n } p_i - T_n)^2,\, \text{s.t. } \eqref{equ: MU Tn and pi}$ and $0\leq p_i \leq 1 $, $\forall i \in \mathcal{I}$.} We can observe that the approximation is accurate with a moderate Zipf exponent. Then, $\lambda_{u,n}^\prime$ can be approximated as
	\begin{equation}\label{equ: MU Appendix lambdaUn App}
	\lambda_{u,n}^\prime \approx { \lambda_{2} U_2  \over C_2} \sum_{i \in \mathcal{I}^n } p_i \overset{(a)}{=} {\lambda_{2} T_n U_2 \over C_2},
	\end{equation}
	where (a) is due to \eqref{equ: MU Tn and pi}.
	
	Next, we will calculate the PMF of $\bar{\varTheta}$. We first give the probability that a user sends a request to an SBS. Consider a user $u_x$ located at $x$, who requests file $n \in \mathcal{N}_c$. This guarantees that $u_x$ only can be served by SBSs. Consider an SBS $B_0$ located at the origin, which stores the file combination $\mathcal{I}_i$. Let $Z_0$ represent the distance between user $u_x$ and its serving SBS. 
	
	Consider the case that $\mu < 1$. When $\|x\| \in [ 0, \mu Z_0]$, $B_0$ will receive an IN request from $u_x$ if it is not the serving SBS of $u_x$, which means $B_0$ does not store file $n$, i.e., $n \notin \mathcal{I}_i$; otherwise, $B_0$ will be the serving SBS of $u_x$, which is contradictory. Therefore, the probability that $u_x$ sends an IN request to $B_0$, denoted by $p_{\text{s}}(x,n | \mathcal{I}_i)$, is given by 
	\begin{align}\label{equ: MU Appendix psend u<1}
		p_{\text{s}}&(x,n | \mathcal{I}_i)  \notag = \mathds{1} \{ n \notin \mathcal{I}_i \}  \mathbb{P} \left[ \| x \| \leq \mu Z_0  \right]   \notag
		\\= & \mathds{1} \{ n \notin \mathcal{I}_i \} \int_{\frac{\|x\|}{\mu}}^{\infty} f_{Z_0}(z) \mathrm{d}z 
		=   \mathds{1} \{ n \notin \mathcal{I}_i \} e^{-\pi \lambda_2 T_n \frac{\|x\|^2}{\mu ^2}  } ,
	\end{align}	
	where $ f_{Z_0}(z) = 2\pi \lambda_2 T_n z e ^{-\pi \lambda_2 T_n z^2}$ is the PDF of the distance between $u_x$ and its serving SBS. Denote by $\Phi_{u,n}^{\prime s}$ the set of served users that request file $n$ and send an IN request to SBS $B_0$, which is a location-dependent thinned PPP of $\Phi_{u,n}^\prime$ with density $\lambda_{u,n}^{\prime s} (x,n,i) = p_{\text{s}}(x,n | \mathcal{I}_i) \lambda_{u,n}^\prime $. Denote by $\varTheta _ {in}$ the number of requests received at $B_0$ from the user in $\Phi_{u,n}^{\prime s}$, which equals the total number of users in $\Phi_{u,n}^{\prime s}$. Therefore, using the Campbell’s Theorem, the mean of $\varTheta _ {in}$ is given by 
	\begin{align}\label{equ: MU Appendix mean of Theta_in mu<1}
		\bar{\varTheta} _ {in} =  &\int_{\mathbb{R}^{2}} \lambda_{u,n}^{\prime s} (x,n,i) \mathrm{d}x \notag
		\\\overset{(a)}{=} & {\lambda_{2} T_n U_2 \over C_2} \!\! \int_{0}^{2\pi} \!\!\!\! \int_{0}^{\infty} \!\! \mathds{1} \{ n \notin \mathcal{I}_i \}  e^{-\pi \lambda_2 T_n \frac{\|x\|^2}{\mu ^2}  } r \mathrm{d}r \mathrm{d} \theta \notag
		\\ = &   \mathds{1} \{ n \notin \mathcal{I}_i \} {U_2 \mu^2 \over C_2},
	\end{align}			
	where (a) is from the polar-Cartesian coordinate transformation, i.e., $\mathrm{d} y  = r \mathrm{d} r \mathrm{d} \theta$. We sum $\bar{\varTheta} _ {in}$, on the right-hand side of \eqref{equ: MU Appendix mean of Theta_in mu<1}, for all the files in $\mathcal{N}_c$: 
	\begin{align}\label{equ: MU Appendix bar varTheta i 1 mu<1}
	\bar{\varTheta} _ {i} = \sum_{n \in \mathcal{N}_c} \bar{\varTheta} _ {in}
	 =  \left( \sum_{n \in \mathcal{N}_c} 1 - \sum_{n \in \mathcal{I}_i} 1 \right)  {U_2 \mu^2 \over C_2}  .
	\end{align}				
	Removing the condition that $B_0$ stores combination $I_i$, we have
	\begin{align}\label{equ: MU Appendix bar varTheta mu<1}
		\bar{\varTheta}  & = \sum_{ i \in \mathcal{I}} p_i \bar{\varTheta} _ {i} = 
		\sum_{ i \in \mathcal{I}} p_i \! \left( \sum_{n \in \mathcal{N}_c} 1 - \sum_{n \in \mathcal{I}_i} 1 \right)  {U_2 \mu^2 \over C_2} \notag
		\\& \overset {(a)}{=} \left( N_c - \sum_{ i \in \mathcal{I}} \sum_{n \in \mathcal{I}_i} p_i \right) {U_2 \mu^2 \over C_2} \overset {(b)}{=}  \left(  N_c -  \sum_{ n \in \mathcal{N}_c}  T_n   \right)   {U_2 \mu^2 \over C_2} \notag
		\\ & =  \left( N_c \! - \! C_2 \right)   {U_2 \mu^2 \over C_2}  ,
	\end{align}				
	where (a) is due to $N_c = \sum_{n \in \mathcal{N}_c} 1$, and (b) is from the relation 	$C_2 = \sum_{i \in \mathcal{I}} p_i C_2 = \sum_{i \in \mathcal{I}}  \sum_{ n \in \mathcal{I}_i} p_i \mathds{1} \{ n \in  \mathcal{I}_i  \}= \sum_{ n \in \mathcal{N}_c}\sum_{ i \in \mathcal{I}^n} p_i =\sum_{ n \in \mathcal{N}_c} T_n $ \cite{Cui2016}.	
	
	When $\mu \geq 1$, we need to consider the following two cases: 1) when $\|x\| \in [ 0,Z_0]$, $B_0$ will receive an IN request from $u_x$ if $n \notin \mathcal{I}_i$; 2) when $ \| x \|  \in (Z_0, \mu Z_0] $, $B_0$ will always receive a request from $u_x$. In this case, similar to \eqref{equ: MU Appendix psend u<1}, $p_{\text{s}}(x,n | \mathcal{I}_i)$ is given by	
	\begin{align}
		p_{\text{s}}&(x,n | \mathcal{I}_i)  \notag = \mathds{1} \{ n \notin \mathcal{I}_i \} \mathbb{P} \left[ \| x \| \leq \mu Z_0  \right] \! + \! \mathbb{P} \left[ Z_0 < \| x \| \leq \mu Z_0  \right] \notag
		\\= &  \left( \mathds{1} \{ n \notin \mathcal{I}_i \} - 1\right) e^{ -\pi \lambda_2 T_n \|x\|^2 } +  e^{-\pi \lambda_2 T_n \frac{\|x\|^2}{\mu ^2} } .
	\end{align}		
	Similarly, $\bar{\varTheta} _ {in}$ can be obtained as 
	\begin{align}\label{equ: MU Appendix mean of Theta_in mu>1}
		\bar{\varTheta} _ {in} \! =\! \int_{\mathbb{R}^{2}} p_{\text{s}}(x,n | \mathcal{I}_i) \lambda_{u,n}^\prime  \mathrm{d}x 
		=   \left(  \mathds{1} \{ n \notin \mathcal{I}_i \} \! - \!1 \right) {U_2 \over C_2}  +  {U_2  \mu^2 \over C_2}  .
	\end{align}		
	After summing $\bar{\varTheta} _ {in}$, on the right-hand side of \eqref{equ: MU Appendix mean of Theta_in mu>1}, for all the files in $\mathcal{N}_c$, and removing the condition that $B_0$ stores combination $I_i$, we have
	\begin{equation}\label{equ: MU Appendix bar varTheta mu>1}
		\bar{\varTheta} =\sum_{ i \in \mathcal{I}} p_i \sum_{n \in \mathcal{N}_c} \bar{\varTheta} _ {in}   = \frac{N_cU_2\mu^2}{C_2} - U_2.
	\end{equation}
	 
	 Combining \eqref{equ: MU Appendix bar varTheta mu<1} and \eqref{equ: MU Appendix bar varTheta mu>1}, we can obtain the expression for $\bar{ \varTheta}$ in \eqref{equ: MU meanTheta}. 

\section{Derivation of $q_2 (\mathcal{N}_c,\mathbf{T},\mu)$}\label{appendix: MU proof Prop q2}
	{\color{black}\subsection{Case $\mu < 1$}
	Based on the total probability formula, we have 
	\begin{equation}\label{equ: MU Appendix Psi2}
		\mathbb{P} \left[ \Upsilon_{n,2}  \geq \tau \right] = \sum_{\theta = 0}^{ \infty }  p_{\varTheta} (\theta)\varPsi_{2,\varTheta }(T_n ,N_c, \mu, \theta),
	\end{equation}
	where $\varPsi_{2, \varTheta }(T_n ,N_c, \mu,\theta) \triangleq \mathbb{P} \left[ \left. \Upsilon_{n,2}  \geq \tau \right| \varTheta = \theta \right]$ is the STP conditioned on the number of received IN requests at the serving SBS of $u_0$ is $\varTheta = \theta$. Conditioning on $Z_2 = z_2$, we have
	\begin{align}\label{equ: MU Appendix Psi2theta}
	\varPsi_{2,\varTheta }(T_n , \! N_c, \! \mu, \! \theta) &  = \!  \!\int_{0}^{\infty}\!\!\! \! \mathbb{P} \left[ \left.\Upsilon_{n,2} \! \geq \! \tau \right| \! Z_2 \!= \!z_2, \! \varTheta  \! =\! \theta \right]  f_{Z_2} (z_2) \mathrm{d}z_2 \notag
	\\& =  \int_{0}^{\infty} \! \!\!\! \mathbb{P}  \left[ g_{02} \geq \tau z_2^{\alpha_2} I \right] f_{Z_2} (z_2) \mathrm{d}z_2,
	\end{align}			
	where $I = \sum_{x \in \tilde{\Phi}_2 } g_{x_2} \|x\|^{-\alpha_2}$, with $\tilde{\Phi}_2 = \Phi_2^a \cup \Phi_2^b \cup \Phi_2^c $; and $f_{Z_2} (z_2) = 2\pi \lambda_2 T_n z_2 \exp (-\pi \lambda_2 T_n z_2^2)$ is the PDF of the distance $Z_2$. Note that applying the IN scheme will only change the density of the interfering SBSs of Type-A (for $\mu<1$) or Type-B (for $\mu \geq 1$). Therefore, the densities of $\Phi_2^a$, $\Phi_2^b$, and $\Phi_2^c$ are $\lambda_{2}^{a} = \varepsilon \lambda_{2 }\left(1-T_{n}\right)$, $\lambda_{2}^{b} = a b \lambda_{2}$, and $\lambda_{2}^{c} = \lambda_{2}$, with $a$ and $b$ given in \eqref{equ: MU a and b}. Since $g_{02} \overset{d}{\sim} \Gamma(D_2, 1)$ where $D_2 = \max \{M_2-U_2+1-\theta , 1\}$ (cf. Section \ref{subsec: MU Signal Model}), using the CDF of $g_{02}$, we have \cite{Li2015a}
	\begin{align}\label{equ: MU Appendix CCDF g02}
		\mathbb{P} \left[ g_{02} \geq \tau z_2^{\alpha_2} I \right] & =  \mathbb{E}_{I } \left[\sum_{m=0}^{D_{2}-1} \frac{s^{m}}{m !}I^{ m} e^{-s I }\right] \notag
		\\& \overset{(a)}{=} \sum_{m=0}^{D_2 -1} \frac{(-s)^{m}}{m !} \mathcal{L}_{I}^{(m)}(s),
	\end{align}	
	where $ s = \tau z_2^{\alpha_2}$; $\mathcal{L}_{I}^{(m)}(s)$ denotes the $m$-th derivative of $\mathcal{L}_{I}(s)$; and (a) is due to the property of the Laplace transform, i.e., $\mathbb{E}_{I} \left[I^{ m} e^{-s I}\right]=(-1)^{m} \mathcal{L}_{ I} ^{(m)} (s)$.
	
	Furthermore, we have	
	\begin{align}\label{equ: MU Appendix Lap_I prime}
		\mathcal{L}_{I} \left( s \right) &=  \mathbb{E}_{I} \left[ \exp ( -s I ) \right]\notag
		\\ & = \mathbb{E}_{I} \!\Bigg[ \exp \Bigg (- s \sum_{x \in \tilde{\Phi}_2 }  g_{x_2} \|x\|^{-\alpha_2} \Bigg ) \Bigg] \notag
		\\& =  \mathbb{E}_{ \tilde{\Phi}_2 } \Bigg[\prod_{x \in \tilde{\Phi}_2 } \mathbb{E}_{ g_{x_2}} \left[ \exp \left( -s g_{x_2} \| x \| ^{-\alpha_2} \right) \right] \Bigg] \notag 
		\\ & \overset{(a)}{=}  \mathbb{E}_{ \tilde{\Phi}_2  } \Bigg[ \prod_{x \in  \tilde{\Phi}_2 } \left( 1 + s \| x \|^{-\alpha_2} \right)^{-U_2} \Bigg] \notag
		\\& \overset{(b)}{=} \exp \Bigg( \!\! \underbrace{ -2\pi \sum_{i = a}^c \lambda_2^i \int_{\varOmega_i} \left( 1- (1 + s v^{-\alpha_2})^{-U_2} \right) v \mathrm{d}v }_{\triangleq \chi (s)} \!\Bigg ),
	\end{align}
	where (a) is due to $ g_{x_2} \overset{d}{\sim} \Gamma (U_2,1)$, (b) is from the probability generating functional for a PPP and the polar-Cartesian coordinate transformation, and the intervals $\varOmega_{a} = [0, \mu z_2]$, $\varOmega_{b} = [\mu z_2, z_2 ]$, and $\varOmega_{c} = [ z_2, + \infty )$ are given in Section \ref{subsec: MU IN scheme}. Denote the exponent in \eqref{equ: MU Appendix Lap_I prime} by $\chi(s)$ as shown above. We have
	\begin{align}\label{equ: MU Appendix chi}
		\chi(s) = & -2\pi ( \lambda_2^a \mathcal{H}_{0}^{\mu z_2} - \lambda_2^b \mathcal{H}_{\mu z_2}^{z_2}  -  \lambda_2^c \mathcal{H}_{ z_2}^{\infty} )	\notag
		\\ =&  -2\pi \left( \lambda_2^a \left( \mathcal{H}_{0}^{\infty}  - \mathcal{H}_{\mu z_2}^{\infty}  \right) \! - \! \lambda_2^b \left( \mathcal{H}_{\mu z_2}^{\infty}  \!- \! \mathcal{H}_{ z_2}^{\infty} \right) \! - \! \lambda_2^c \mathcal{H}_{z_2}^{\infty}	 \right)	\notag						
		\\= & -  \pi \lambda_2^a \left(  G \left(s \right) - (\mu z_2)^2 F_2(s (\mu z_2)^{-\alpha_2}) \right) \notag
		\\& - \pi \lambda_2^b \left( (\mu z_2)^2 F_2(s (\mu z_2)^{-\alpha_2})  - z_2^2 F_2(s z_2^{-\alpha_2}) \right) \notag
		\\ &-  \pi \lambda_2^c  z_2^2 F_2(s z_2^{-\alpha_2}),
	\end{align}
	 where $ \mathcal{H}_{x}^{y} = \int_{x}^{y} h(v)\mathrm{d}v $, with $h(v) \triangleq \left( 1- (1 + s v^{-\alpha_2})^{-U_2} \right) v$; and $G(x)$ is given by \eqref{equ: MU Gj}. In \eqref{equ: MU Appendix chi}, $\mathcal{H}_{x}^{\infty}$ can be obtained from \eqref{equ: MU Appendix LI1} when $x>0$, whereas $\mathcal{H}_{0}^{\infty}$ can be obtained by $\lim\limits_{z \rightarrow 0} z^2 F_2(s z ^{-\alpha_2}) =  G(s)$.
	Considering $\mathcal{L}_{I} ^ \prime (s)= \mathcal{L}_{I} (s) \chi ^ \prime (s)$ and the Leibniz formula, we have
	\begin{equation}\label{equ: MU Appendix Leibniz}
	\mathcal{L}_{I}^{(m)}(s)=\sum_{k=0}^{m-1}\binom{m-1}{k} \chi^{(m-k)}(s) \mathcal{L}_{I}^{(k)}(s),
	\end{equation}	
	where $\chi^{(k)}(s)$ denotes the $k$-th derivative of $\chi(s)$, and is given by 
	\begin{align}\label{equ: MU Appendix chi_k}
		\chi&^{(k)}(s) \notag
		\\= &  2\pi \sum_{i = a}^c \lambda_2^i \int_{\varOmega_i} (-1)^k (U_2)_k  \frac{\left( v^{-\alpha_2} \right)^k v  }{ \left( 1 + s v^{-\alpha_2} \right)^{U_2 + k} }\mathrm{d}v \notag
		\\= & (-1)^k (U_2)_k 2\pi \Bigg\{  \lambda_2^a \! \bigg[ \frac{1}{\alpha_2} s^{\frac{2}{\alpha_2} -k}  B \left(k -  \frac{2}{\alpha_2}, U_2  +  \frac{2}{\alpha_2}\right)  \notag
		\\& -   (\mu z_2) ^{2 }    \tilde{F}_{k} (   s (\mu z_2)^{-\alpha_2} \! ) \bigg]    +   \lambda_2^b \bigg[ (\mu z_2) ^{2}   \tilde{F}_{k}  (s (\mu z_2)^{ -\alpha_2} ) 	 \notag 
		\\&- z_2^ {2} \tilde{F}_{k}(s z_2^{-\alpha_2} )  \bigg] 
		+ \lambda_2^c z_2^{2} \tilde{F}_{k}(s z_2^{-\alpha_2} )   \Bigg\},
	\end{align}			
	where $(U)_n = U(U+1)\cdots(U+n-1)$ is the Pochhammer symbol, $B(x,y)$ is the Beta function, and $\tilde{F}_{k}(x)$ is given in \eqref{equ: MU Fjk}. The second equality above is due to \cite[(3.194)]{Zwillinger2014}. Let 
	\begin{equation}
		w_0 = -\frac{\chi(s)}{ \pi z_2^{2} \lambda_2 } , \quad w_m = \frac{(-s)^{m}}{\pi z_2^{2} \lambda_2 m !} \chi^{(m)}(s),
	\end{equation}
	we thus have $\left. \mathcal{L}_{I}(s)\right|_{s = \tau z_2^{\alpha_2}} = \exp (-\pi z_2^{2} w_0 ) $. Substituting $s = \tau z_2^{\alpha_2}$ into \eqref{equ: MU Appendix chi} and \eqref{equ: MU Appendix chi_k}, after some manipulation, the expressions of $w_0$ and $w_m$ are obtained as \eqref{equ: MU w0} and \eqref{equ: MU wm}.
	
	Let $x_{m}=\frac{1}{m !}(-s)^{m} \mathcal{L}_{I}^{(m)}(s)$. From \eqref{equ: MU Appendix Leibniz}, we have
	\begin{equation} \label{equ: MU Appendix xm}
		x_{m}=\sum_{k=0}^{m-1} \frac{m-k}{m}\left(\frac{(-s)^{m-k}}{(m-k) !} \chi^{(m-k)}(s)\right) x_{k}.
	\end{equation}
	Considering the recurrence relation in \eqref{equ: MU Appendix xm}, we have		
	\begin{equation}\label{equ: MU Appendix xm wm}
		x_{m}= \pi z_2^{2} \sum_{k=0}^{m-1} \frac{m-k}{m} w_{m-k} x_{k}, \,\, x_0 = \exp (-\pi z_2^{2} w_0).
	\end{equation}	
	Next, we will obtain an explicit expression for $x_m$ by solving the recurrent relation in \eqref{equ: MU Appendix xm wm}. We first define two power series $\tilde{ W }(z)$ and $X(z)$ as
	\begin{equation}\label{equ: MU Appendix Define of P X}
		\tilde{ W } (z) \triangleq \sum_{m=0}^{\infty} w_{m}  z^{m}, \quad X(z) \triangleq \sum_{m=0}^{\infty} x_{m} z^{m}.
	\end{equation}	
	The derivative of $\tilde{ W }(z)$ is $\tilde{ W }^{(1)}(z)=\sum_{m=0}^{\infty} m w_{m}  z^{m-1}$, the derivative of $X(z)$ is $X^{(1)}(z)=\sum_{m=0}^{\infty} m x_{m}  z^{m-1}$, and the product of $\tilde{ W }(z)$ and $X(z)$ is $\tilde{ W }(z) X(z)=\sum_{m=0}^{\infty} (\sum_{i=0}^{m}  w_{m-i}  x_{i} ) z^{m}$. Then, from \eqref{equ: MU Appendix xm wm}, we have
	\begin{equation}\label{equ: MU Appendix diff equ2}
		zX^{(1)}(z) = \pi z_2^{2} z \tilde{ W }^{(1)}(z) X(z).
	\end{equation}
	By solving the above differential equation, we have $X(z)= C \exp \left( \pi z_2^{2} \tilde{ W }(z)\right)$, where $C$ is a constant to be determined. Since $X(0) = x_0 = \mathcal{L}_{I} (s)$ and $\tilde{ W }(0) = w_0 $, from \eqref{equ: MU Appendix xm wm}, we can obtain $C = \exp (- 2 \pi z_2^{2} w_0 )$, and
	\begin{equation}\label{equ: MU appendix solution of X}
		X(z) = \exp \left( \pi z_2^{2} \left(   \tilde{ W }(z) -2w_0 \right) \right).
	\end{equation}
	
	Recalling from \eqref{equ: MU Appendix Psi2theta}, we have 
	\begin{align}\label{equ: MU Appendix Psi2Theta_FinalExpression}
		\varPsi&_{2,\varTheta } (T_n , N_c,  \mu,  \theta) \notag
		\\= & \mathbb{E}_{Z_2} \left[  \mathbb{P} \left[ g_{02} \geq \tau z_2^{\alpha_2} I \right] \right] = \mathbb{E}_{Z_2} \left[ \sum_{m=0}^{D_2 -1} x_m \right] \notag
		\\ = & \mathbb{E}_{Z_2} \left[ \sum_{m=0}^{D_2 -1} \left. \frac{1}{m!} \frac{\mathrm{d}^{m}}{\mathrm{d} z^{m}} X(z)\right|_{z=0} \right] \notag
		\\ = & \sum_{m=0}^{D_2 -1} \left. \frac{1}{m!} \frac{\mathrm{d}^{m}}{\mathrm{d} z^{m}} \mathbb{E}_{Z_2} \left[ X(z) \right] \right|_{z=0} \notag
		\\ = &  \sum_{m=0}^{D_2 -1} \left. \frac{1}{m!} \frac{\mathrm{d}^{m}}{\mathrm{d} z^{m}} \! \int_{0}^{\infty} \!\! \! \! \exp \! \left( \pi z_2^{2} \left(   \tilde{ W }(z) \!- \! 2w_0 \right) \! \right) \! f_{Z_2} (z_2) \mathrm{d}z_2 \right|_{z=0} \notag
		\\ = & { T_n  } \sum_{m=0}^{D_{2}-1} \left.\frac{1}{m !} \frac{\mathrm{d}^{m}}{\mathrm{d} z^{m}} \left( T_n + 2 w_0  - \tilde{ W } (z) \right) ^{-1}  \right|_{z=0} \notag
		\\ \overset{(a)}{=} & T_n  \left\| \left[  \left(  T_n + 2 w_0  \right) \mathbf{I}_{D_2} - \tilde{ \mathbf{W} } _{D_2}(T_n, N_c, \mu) \right] ^{-1} \right\|_1,
	\end{align}
	where $D_2 = \max \{ M_2-U_2+1-\theta , 1 \}$, and $\mathbf{W}_{D_2}(T_n, N_c, \mu) \triangleq \left(  T_n + 2 w_0  \right) \mathbf{I}_{D_2} - \tilde{ \mathbf{W} } _{D_2}(T_n, N_c, \mu) $ is given in \eqref{equ: MU matrixW2}.				
	Let $Q(z) \triangleq T_n (  T_n + 2 w_0 - \tilde{ W }(z) ) ^{-1}$, from \cite[p.~14]{henrici1993applied}, (a) is due to that the first $D_2$ coefficients of $Q(z)$ is the first column of the matrix $T_n  \mathbf{W}_{D_2}(T_n, N_c, \mu) ^{-1} $, and their sum equals the $L_1$ induced norm of this matrix. Further considering \eqref{equ: MU Appendix Psi2}, we can obtain $ \varPsi_{2} (T_n,N_c, \mu)  $ as in \eqref{equ: MU Psi2}, where the second term is due to the fact $\sum_{\theta = M_2-U_2  }^{ \infty } p_{\varTheta} (\theta)  = \frac{\gamma( M_2-U_2, \bar{\varTheta}) }{\Gamma (M_2-U_2 )} $ for the Poisson random variable $\varTheta$. 
	
	\subsection{Case $\mu \geq 1$}
	When $\mu \geq 1$, the integration intervals in \eqref{equ: MU Appendix Lap_I prime},  \eqref{equ: MU Appendix chi}, and \eqref{equ: MU Appendix chi_k} become $\varOmega_{a} = [0,  z_2]$, $\varOmega_{b} = [z_2, \mu z_2 ]$, and $\varOmega_{c} = [ \mu z_2, + \infty )$. Then, following the same procedure as the case $\mu < 1$, we can complete the proof of the case $\mu \geq 1$. We omit the detailed steps due to the page limitation.}

\section{Derivation of the Lower Bound on $\varPsi_{2}(T_n, N_c, \mu)$}\label{appendix: MU proof theorem Two Bounds}
	We define $\mathbf{p} \triangleq [1,1,\cdots,1]_{D_2\times 1}^{T}$ and $\mathbf{q} \triangleq \mathbf{W}_{D_{2}}(T_n, N_c, \mu) \mathbf{p}  $, with $\mathbf{W}_{D_{2}}(T_n, N_c, \mu)$ given in \eqref{equ: MU matrixW2}. Then, we have $\mathbf{p} = \mathbf{W}_{D_{2}}(T_n, N_c, \mu)^{-1}  \mathbf{q} $. Considering the inequality $\|\mathbf{p}\|_{1} \leq \left\|\mathbf{W}_{D_{2}}(T_n, N_c, \mu)^{-1} \right\|_{1} \| \mathbf{q} \|_{1}$, we have 
	\begin{equation}
		\left\|\mathbf{W}_{D_{2}}(T_n, N_c, \mu)^{-1} \right\|_{1} \geq {\|\mathbf{p}\|_{1} \over \| \mathbf{q} \|_{1} }.
	\end{equation}
	Since $\|\mathbf{p}\|_{1} = D_2 $ and $ \| \mathbf{q} \|_{1} =   T_n + w_0  + \sum_{m=1}^{D_2 - 1} [ T_n + w_0 - ( D_2 - m ) w_m]$, we can obtain the following lower bound on $\left\|\mathbf{W}_{D_{2}}(T_n, N_c, \mu)^{-1} \right\|_{1} $:
	\begin{equation}\label{equ: MU Appendix lowerbound on W-1}
		\left\|\mathbf{W}_{D_{2}}(T_n, N_c, \mu)^{-1} \right\|_{1} \geq { 1 \over T_n + w_0 - \sum_{m=1}^{D_2 - 1 } \left( 1 - { m \over D_2} \right) w_m }.
	\end{equation}
	Replacing $\left\|\mathbf{W}_{D_{2}}(T_n, N_c, \mu)^{-1} \right\|_{1}$ in \eqref{equ: MU Psi2} with the right-hand side term in \eqref{equ: MU Appendix lowerbound on W-1}, and after some manipulation, we can obtain the lower bound $\varPsi_{2}^l  (T_n, N_c, \mu)$ shown in \eqref{equ: MU Psi2 lower}.
	
\section{Proof of Property \ref{proper: MU Nb consecutive}}\label{appendix: MU proof proper Nb consec}
	Let $( \mathcal{N}_c^\star, \mathbf{T}^\star )$ and $\mathcal{N}_b^\star =  \mathcal{N}_2 \backslash \mathcal{N}_c^\star $ denote an optimal solution to Problem \ref{Prob: MU Cache Placement}. Suppose $\mathcal{N}_b^{ \star}$ has $S$ elements and is represented by $\mathcal{N}_b^{ \star} = \{ n_1^\star,n_2^\star,  \cdots, n_{S}^\star \}$, the elements of which are non-consecutive and sorted in ascending order. Note that the non-consecutiveness of the elements in $\mathcal{N}_b^{ \star} $ is equivalent to $n_{S}^\star  - n_1^\star + 1 >S $, which is also equivalent to the existence of an index $j$, $1\leq j < S$, such that $n_j^\star +1 < n_{j+1}^\star $. We now construct from $( \mathcal{N}_c^\star, \mathbf{T}^\star )$ another optimal solution, where the indexes of files in $\mathcal{N}_b^\star $ are consecutive. 
		
	Denote by $f(x,y) \triangleq \min \{ 1, {C_b \over N_2 - y} \}  \lambda_{1} U_1 \varPsi_{1} - \lambda_{2} U_2 \varPsi_{2}( x , y, \mu )$. We first show that $  f( T_{ n_j^\star+1 }^\star, N_c^\star )  = 0 $.
	
	Suppose $  f( T_{ n_j^\star+1 }^\star, N_c^\star )  < 0 $. Consider the alternative solution $\mathcal{N}_{b}^{\text{up}} =\! \{ n_2^\star,\cdots, n_j^\star, n_j^\star+1, n_{j+1}^\star,  \cdots, n_{S}^\star  \}$, $T_{  n_1^{ \star}  }^{\text{up}}= T_{ n_j^\star+1 }^\star$, and $T_n^{\text{up}} = T_n^\star$ for $\forall n \in \mathcal{N}_c^{\star} \backslash  \{  n_j^\star+1 \}  $. Then, we have $ \mathrm{ASE} ( \mathcal{N}_c^{\text{up}}, \mathbf{T}^{\text{up}} ,\mu ) - \mathrm{ASE}(\mathcal{N}_c^\star, \mathbf{T}^\star, \mu) =   \log_{2}(1+\tau) (  a_{ n_j^\star+1 } - a_{ n_1^\star }  )  f( T_{ n_j^\star+1 }^\star, N_c^\star )  $. Since $n_1^\star < n_{n_j^\star+1}$, we have $a_{ n_1^\star } >  a_{ n_j^\star+1}$, and thus $ \mathrm{ASE} ( \mathcal{N}_c^{\text{up}}, \mathbf{T}^{\text{up}} ,\mu  ) - \mathrm{ASE}(\mathcal{N}_c^\star, \mathbf{T}^\star, \mu ) > 0$, which contradicts with the optimality of $( \mathcal{N}_c^\star, \mathbf{T}^\star )$.
	
	Suppose $  f( T_{ n_j^\star+1 }^\star, N_c^\star )  > 0 $. Consider the alternative solution $\mathcal{N}_{b}^{\text{dn}} =\! \{n_1^\star,\cdots, n_j^\star, n_j^\star+1, n_{j+1}^\star,  \cdots, n_{S-1}^\star  \}$, $T_{  n_S^{ \star}  }^{\text{dn}} = T_{ n_j^\star+1 }^\star$, and $T_n^{\text{dn}} = T_n^\star$ for $\forall n \in \mathcal{N}_c^{\star} \backslash  \{  n_j^\star+1 \}  $. Then, we have $ \mathrm{ASE} ( \mathcal{N}_c^{\text{dn}}, \mathbf{T}^{\text{dn}} ,\mu ) - \mathrm{ASE}(\mathcal{N}_c^\star, \mathbf{T}^\star, \mu) =   \log_{2}(1+\tau) (  a_{ n_j^\star+1 } - a_{ n_S^\star }  )  f( T_{ n_j^\star+1 }^\star, N_c^\star )  $. Similarly to the previous case, we can show that this is strictly greater than zero, leading to contradiction.
	
	Now, since $  f( T_{ n_j^\star+1 }^\star, N_c^\star )  = 0 $, we can construct a new solution $\mathcal{N}^\prime = \{n_1^\prime, n_2^\prime, \cdots, n_S^\prime \}$ as equal to either $\mathcal{N}_{b}^{\text{up}} $ or $\mathcal{N}_{b}^{\text{dn}}$, and we have $ \mathrm{ASE} ( \mathcal{N}_c^\prime, \mathbf{T}^\prime ,\mu ) - \mathrm{ASE}(\mathcal{N}_c^\star, \mathbf{T}^\star, \mu) =0$, which means $( \mathcal{N}_c^\prime, \mathbf{T}^\prime )$ is also optimal. Note that in ether case, we have $n_{S}^\prime  - n_1^\prime + 1 < n_{S}^\star  - n_1^\star + 1 $.

	 Relabel the elements in $\mathcal{N}_b^\prime $ as $\{ n_1^\star, \cdots, n_{S}^\star  \}$, and relabel $\mathcal{N}_b^\prime $ as $\mathcal{N}_b^\star$. Note that by doing so, we have decreased the value of $n_{S}^\star  - n_1^\star + 1$ without losing optimality. If the elements of $\mathcal{N}_b^\star$ are still non-consecutive, we can repeat the above procedure to decrease the value of $n_{S}^\star  - n_1^\star + 1$ each time, until it reaches $n_{S}^\star  - n_1^\star + 1 = S$. At this point, the elements of $\mathcal{N}_b^\star$ are consecutive (and $\mathcal{N}_b^\star$ is still optimal).

\section{Solution to Problem \ref{Prob: MU Lowerbound Problem}}\label{appendix: MU KKT}	
The Lagrange function of Problem~\ref{Prob: MU Lowerbound Problem} is given by
	\begin{equation}
	\begin{aligned}
		\mathcal{L} =& -\mathrm{ASE}^l (\mathcal{N}_c,\mathbf{T},\mu)  - \sum_{n \in \mathcal{N}_c} \lambda_{n} T_{n} \\&+\sum_{n \in \mathcal{N}_c} \vartheta_{n} \left( T_{n}-1 \right) +\nu \left(\sum_{n \in \mathcal{N}_c} T_{n} - C_2 \right),
	\end{aligned}	
	\end{equation}
	where $ \boldsymbol{\lambda} \triangleq \left[ \lambda_{n} \right] _ { n \in \mathcal{N}_c}$ and $\boldsymbol{\vartheta} \triangleq \left[ \vartheta_{n} \right] _{n \in \mathcal{N}_c}$ are the Lagrange multipliers associated with the constraint $0\leq T_n \leq 1$, $\forall n \in \mathcal{N}_c$; and $\nu$ is the Lagrange multiplier associated with the constraint $\sum_{n \in \mathcal{N}_c}  T_{n} = C_2$. Then, the derivative of $\mathcal{L}$ w.r.t. $T_n$ is 
	\begin{align}\label{equ: MU Appendix DL}
		\frac{\partial \mathcal{L}  }{\partial T_n} = & - a_n   f (T_n) - \lambda_n + \vartheta_n + \nu,
	\end{align}		
	where $f(x) $ is given in \eqref{equ: MU derivative f}.
	
	The KKT conditions can be written as
	\begin{subequations}
		\begin{align}
			0 \leq T_{n}^\dagger \leq 1, \quad \sum\nolimits_{n \in \mathcal{N}_c} T_{n}^\dagger &= C_2, \label{equ: MU KKT 1}\\
			\lambda_n^\dagger \geq 0,\quad \vartheta_n^\dagger &\geq 0, \label{equ: MU KKT 2}\\
			\lambda_n^\dagger T_n^\dagger = 0,\quad \vartheta_n^\dagger (T_n^\dagger-1) &= 0,\label{equ: MU KKT 3} \\
			\left. \frac{\partial \mathcal{L}  }{\partial T_n}\right|_{T_n = T_n^\dagger} &= 0, \label{equ: MU KKT 4}
		\end{align}
	\end{subequations}
	where $\lambda_n^\dagger$ and $\vartheta_n^\dagger$, for $\forall n \in \mathcal{N}_c$, are the optimal value of $\lambda_n$ and $\vartheta_n$, respectively.
	From \eqref{equ: MU KKT 4}, we have $\vartheta_n^\dagger = a_n f(T_n^\dagger) + \lambda_n^\dagger - \nu^\dagger $. Since $f(x)$ is a decreasing function of $x$, we have the following three cases: If $ \nu^\dagger > a_n f_{\max}(x) = a_n f(0)$, considering the dual feasibility in \eqref{equ: MU KKT 2}, the complementary slackness condition in \eqref{equ: MU KKT 3} can only hold if $ \lambda_n^\dagger > 0 $ and $ \vartheta_n^\dagger =0 $, implying that $ T_n^\dagger = 0 $. If $ \nu^\dagger < a_n f_{\min}(x) = a_n f(1)$, the conditions in \eqref{equ: MU KKT 2} and \eqref{equ: MU KKT 3} yield $ \lambda_n^\dagger = 0 $ and $ T_n^\dagger = 1 $. If $a_n f(1) \leq  \nu^\dagger \leq a_n f(0)$, \eqref{equ: MU KKT 2} and \eqref{equ: MU KKT 3} can only hold if $\lambda_{n}^\dagger = \vartheta_{n}^\dagger = 0$, so from \eqref{equ: MU Appendix DL} and \eqref{equ: MU KKT 4}, we have $ \nu_n^\dagger = a_n f(T_n^\dagger)$. By solving the equation $ \nu_n^\dagger = a_n f(T_n^\dagger)$ as well as \eqref{equ: MU KKT 1}, we can obtain $T_n^\dagger $ and its corresponding $\nu_n^\dagger$.
	
\section{CCP Approach for Maximizing $\mathrm{ASE}^u (\mathcal{N}_c, \mathbf{T}, \mu)$ for the Problem in \eqref{equ: MU Prob Equivalent 2} }\label{appendix: MU CCP}	
	\subsection{An Upper Bound on $ \mathrm{ASE} (\mathcal{N}_c,\mathbf{T},\mu)$}
	From Appendix \ref{appendix: MU proof Prop q2}, we have
	\begin{align}
	\varPsi_{2,\varTheta }
	& = \mathbb{E}_{I,Z_2} \left[  \mathbb{P} \left[ g_{02} \geq \tau z_2^{\alpha_2} I \right] \right] \notag
	\\&\stackrel{(a)}{\leq} 1- \mathbb{E}_{I,Z_2} \left[ \left( 1- \exp \left(-\beta \tau z_2^{\alpha_2} I\right) \right)^{D_2} \right] \notag
	\\& = 1- \mathbb{E}_{I,Z_2} \left[ \sum_{i=0}^{D_2} (-1)^{i} \binom{D_2}{i} \exp \left(- i \beta \tau z_2^{\alpha_2} I\right)   \right] \notag
	\\& = \sum_{i=1}^{D_2} (-1)^{i+1} \binom{D_2}{i} \mathbb{E}_{I,Z_2} \left[ \exp \left(- i \beta \tau z_2^{\alpha_2} I\right) \right] \notag
	\\&=\sum_{i=1}^{D_2} (-1)^{i+1} \binom{D_2}{i}  \mathbb{E}_{Z_2} \left[ \mathcal{L}_{I} \left( i \beta \tau z_2^{\alpha_2} \right) \right]   \triangleq \varPsi_{2,\varTheta }^u ,
	\end{align}		
	where (a) is due to $g_{02} \overset{d}{\sim} \Gamma(D_2, 1)$ and a lower bound on the incomplete gamma function ${\gamma(D,x) \over \Gamma(D)}$, i.e., $\left( 1- e^{-\beta x} \right) ^D \leq {\gamma(D,x) \over \Gamma(D)}$ \cite{Horst1997}. Since $ \mathcal{L}_{I}( \tau z_2^{\alpha_2}) = \exp (-\pi z_2^{2} w_0 ) $, we can obtain $\mathcal{L}_{I} \left( i \beta \tau z_2^{\alpha_2} \right)$ by replacing $\tau$ with $i\beta \tau$, i.e., $ \mathcal{L}_{I} \left( i \beta \tau z_2^{\alpha_2} \right) = \exp \left( -\pi z_2^2 A_i \right)$, with $A_i \triangleq  \varepsilon \left(1-T_{n}\right) G \left(  i \beta \tau  \right) +   a  ( 1 - \varepsilon ) \mu^2   F_2\left({  i \beta \tau \over \mu ^{ \alpha_2 }} \right) + b T_n F_2(  i \beta \tau )$, $\beta = (D_2 !)^{-1/D_2 }$. We then have $ \mathbb{E}_{Z_2} \left[ \mathcal{L}_{I} \left( i \beta \tau z_2^{\alpha_2} \right) \right] = \int_{0}^{\infty}\exp \left( -\pi z_2^2 A_i   \right)  f_{Z_2} (z_2) \mathrm{d}z_2 = { T_n \over  \varrho_{1,i} ( \beta ) T_n +   \varrho_{2,i} ( \beta )  } $, with $\varrho_{1,i}(x) $ and $\varrho_{2,i}(x) $ defined in \eqref{equ: MU varrho1ix} and \eqref{equ: MU varrho2ix}. Note that $ \varPsi_{2}^{u} (T_n, N_c, \mu) = \sum_{\theta = 0}^{ \infty }  p_{\varTheta} (\theta) \varPsi_{2,\varTheta } ^{u} $. After some manipulation, we obtain an upper bound on $\varPsi_{2}(T_n, N_c, \mu)$ given by
	\begin{align}
		\varPsi_{2}^u & (T_n, N_c, \mu) \notag
		\\ = &  \sum_{\theta = 0}^{  M_2-U_2 - 1 } p_{\varTheta} (\theta)  \sum_{i=1}^{D_2} (-1)^{i+1} \binom{D_2}{i} { T_n \over  \varrho_{1,i} ( \beta ) T_n +   \varrho_{2,i} ( \beta )  } \notag
		\\ &  + { p_{\varTheta}^s   T_n \over \varrho_{1,1} ( 1 ) T_n +   \varrho_{2,1} ( 1 ) } ,\label{equ: MU Psi2 upper}
	\end{align}
	where $\varrho_{1,i}(x) $ and $\varrho_{2,i}(x) $ are given in \eqref{equ: MU varrho1ix} and \eqref{equ: MU varrho2ix}. By replacing the $\varPsi_{2}(T_n, N_c, \mu)$ with $\varPsi_{2} ^l (T_n, N_c, \mu)$ in \eqref{equ: MU q2}, we can obtain an upper bound on the ASE, denoted by $ \mathrm{ASE}^u (\mathcal{N}_c,\mathbf{T},\mu)$. 
	\subsection{CCP Approach for Maximizing $ \mathrm{ASE}^u (\mathcal{N}_c,\mathbf{T},\mu) $}
	When setting $ \mathrm{ASE}^u (\mathcal{N}_c,\mathbf{T},\mu) $ as the objective function of Problem~\ref{Prob: MU Original Problem}, the objective function in \eqref{equ: MU Prob Equivalent 2} will be replaced by its corresponding upper bound, denoted by $\mathrm{ASE}_{2}^{u} (\mathcal{N}_c, \mathbf{T}, \mu )$. It is easy to see that we can decompose $\mathrm{ASE}_{2}^{u} (\mathcal{N}_c, \mathbf{T}, \mu ) $ as
	\begin{equation}
	\mathrm{ASE}_{2}^{u} (\mathcal{N}_c,\mathbf{T},\mu ) =  \eta_1 ( \mathcal{N}_c,\mathbf{T},\mu ) - \eta_2 ( \mathcal{N}_c,\mathbf{T},\mu ),
	\end{equation}
	with
	\begin{align}
	\eta_1 ( \mathcal{N}_c,& \mathbf{T},\mu )   \triangleq \lambda_{2} U_2 \log_{2} ( 1 \! + \! \tau)  \!\! \sum_{n \in \mathcal{N}_c} \! a_n \! \bigg[ { p_{\varTheta}^s   T_n \over \varrho_{1,1} ( 1 ) T_n +   \varrho_{2,1} ( 1 ) } \notag
	\\& + \sum_{\theta = 0}^{  M_2-U_2-1 }  \!\! p_{\varTheta} (\theta)  \sum_{i \in \mathcal{D}_o } \! \! \binom{D_2}{i} { T_n \over  \varrho_{1, i} ( \beta ) T_n +   \varrho_{2, i} ( \beta )  }  \bigg], \notag
	\\\eta_2 ( \mathcal{N}_c,& \!\mathbf{T},\mu ) \!\triangleq \! \lambda_{2} U_2 \log_{2} ( 1 \! + \! \tau) \sum_{n \in \mathcal{N}_c} a_n  \sum_{\theta = 0}^{  M_2-U_2 - 1 } p_{\varTheta} (\theta) \notag
	\\& \times \sum_{i \in \mathcal{D}_e } \! \binom{D_2}{i} { T_n \over  \varrho_{1, i} ( \beta ) T_n +   \varrho_{2, i} ( \beta )  },
	\end{align}
	where $ \mathcal{D}_o$ and $\mathcal{D}_e$ denote the sets of all the odd numbers and even numbers in the set $ \{1,2,\cdots, D_2\}$. Therefore, the continuous optimization problem can be reformulated as follows. 	
	\begin{problem}[Upper Bound Optimization]\label{Prob: MU Upperbound Problem}		
		\begin{align}
		\underset{\mathbf{T} } {\max} \,\,  &\eta_1 ( \mathcal{N}_c,\mathbf{T},\mu ) - \eta_2 ( \mathcal{N}_c,\mathbf{T},\mu )  \notag \\
		&\text{s.t.} \quad \eqref{equ: MU Prob Con2}. \label{equ: MU Prob Upperbound}
		\end{align}		
	\end{problem}
	
	Note that $\eta_1 ( \mathcal{N}_c, \mathbf{T},\mu )  $ and $\eta_2 ( \mathcal{N}_c, \mathbf{T},\mu )  $ are differentiable and concave w.r.t. $\mathbf{T}$. Therefore, Problem \ref{Prob: MU Upperbound Problem} is a DC programming problem, a stationary point of which can be found by using the CCP approach \cite{Lipp2016}. Under CCP, $ - \eta_2 ( \mathcal{N}_c, \mathbf{T}, \mu )$ is linearized using first-order Taylor expansion around the point obtained in the previous iteration, so that the objective function becomes concave and the linearized problem is convex. CCP proceeds by solving such a sequence of convex problems.
	
	Specifically, at the $j$-th iteration, we consider the following problem:
	\begin{align}
	\mathbf{T}^{(j)\dagger} = \arg \underset{\mathbf{T} } {\max} \,\,  &\eta_1 ( \mathcal{N}_c,\mathbf{T},\mu ) - \tilde{ \eta}_2 ( \mathcal{N}_c,\mathbf{T},\mu| \mathbf{T}^{ (j-1) \dagger } )  \notag \\
	&\text{s.t.} \quad \eqref{equ: MU Prob Con2}, \label{equ: MU Prob CCP}
	\end{align}	
	where $\mathbf{T}^{(j)\dagger}$ denotes the optimal value of $\mathbf{T}$ at the $j$-th iteration; and $\tilde{ \eta}_2 ( \mathcal{N}_c,\mathbf{T},\mu| \mathbf{T}^{ (j-1) \dagger } ) \triangleq \eta_2 ( \mathcal{N}_c, \mathbf{T}^{(j-1) \dagger } ,\mu ) + ( \mathbf{T} - \mathbf{T}^{(j-1) \dagger} )^T \nabla  \eta_2 ( \mathcal{N}_c, \mathbf{T}^{(j-1) \dagger} ,\mu ) $, with $\nabla  \eta_2 ( \mathcal{N}_c, \mathbf{T} ,\mu ) $ being the gradient of $ \eta_2 ( \mathcal{N}_c, \mathbf{T} ,\mu ) $ w.r.t. $\mathbf{T} $.
	
	It can be easily observed that the optimization in \eqref{equ: MU Prob CCP} is a convex problem. Similar to the procedures shown in Appendix~\ref{appendix: MU KKT}, by using the KKT conditions, we can obtain the optimal solution to the problem in \eqref{equ: MU Prob CCP} as
	
	\begin{equation} \label{equ: MU Solution to CCP}
	T_{n}^{(j) \dagger}=\left\{\begin{array}{ll}{0,} & {a_n \tilde{f}(0) < \nu^\dagger  ,} \\ {1,} & { a_n \tilde{f}(1) >  \nu^\dagger ,} \\ {x( T_{n}^{(j-1) \dagger}, \nu^\dagger ),} & {\text { otherwise, }}\end{array}\right.
	\end{equation}
	where $\tilde{f}(x) \triangleq \eta_1^\prime ( \mathcal{N}_c, x ,\mu ) - \eta_2^\prime ( \mathcal{N}_c, T_{n}^{(j-1) \dagger} ,\mu )$, with
	\begin{align}
	\eta_1^\prime ( \mathcal{N}_c,& x ,\mu )  \triangleq  \left.{  \partial  \eta_1 ( \mathcal{N}_c, \mathbf{T} ,\mu ) \over a_n \partial T_n } \right|_{T_n = x} =  \lambda_{2} U_2 \log_{2} ( 1 + \tau) \notag
	\\ & \times \bigg[ { p_{\varTheta}^s  \varrho_{2,1} ( 1 )  \over ( \varrho_{1,1} ( 1 ) x +   \varrho_{2,1} ( 1 ) )^2 } + \sum_{\theta = 0}^{  M_2-U_2 -1} p_{\varTheta} (\theta )  \notag
	\\ & \times \sum_{i \in \mathcal{D}_o }   \binom{D_2}{i} {  \varrho_{2, i} ( \beta ) \over  ( \varrho_{1, i} ( \beta ) x +   \varrho_{2, i} ( \beta ) )^2 }  \bigg], \notag
	\\\eta_2^\prime ( \mathcal{N}_c,& x ,\mu )  \triangleq  \left.{  \partial  \eta_2 ( \mathcal{N}_c, \mathbf{T} ,\mu ) \over a_n \partial T_n } \right|_{T_n = x} \!\!=\!\!  \lambda_{2} U_2 \log_{2} ( 1 + \tau) \notag
	\\&  \times \!\! \!\! \sum_{\theta = 0}^{  M_2-U_2-1 } \!\! \!\! p_{\varTheta} (\theta ) \!\!  \sum_{ i \in \mathcal{D}_e }  \!\!\binom{D_2}{i} { \varrho_{2, i} ( \beta )  \over ( \varrho_{1, i} ( \beta ) x \! + \!  \varrho_{2, i} ( \beta ) )^2  }  ,	
	\end{align}	
	where $x( T_{n}^{(j-1) \dagger}, \nu^\dagger )$ denotes the root of equation $a_n \tilde{f}( x ) = \nu^\dagger  $, and the optimal Lagrangian multiplier $\nu^\dagger $ satisfies $\sum_{n \in \mathcal{N}_c} T_{n}^{(j) \dagger} ( \nu ) = C_2$.

	Thus, following the framework of CCP, by iteratively solving the optimization problem in \eqref{equ: MU Prob CCP} according to \eqref{equ: MU Solution to CCP}, a stationary point of Problem \ref{Prob: MU Upperbound Problem} can be obtained. By following the similar procedures introduced in Secton~\ref{subsection: MU Alternating Opt}, we can finally reach a stationary point for Problem~\ref{Prob: MU Original Problem}, when taking the upper bound as the objective function.

	
	\ifCLASSOPTIONcaptionsoff
	\newpage
	\fi
	\bibliographystyle{IEEEtran}
	\bibliography{IEEEabrv,mylib_abrv}

\vfill
\end{document}